\documentclass[12pt,reqno]{amsart}
\allowdisplaybreaks[4]
\usepackage{graphicx} %We can use any other package if it is necessary
\usepackage{amssymb}
\usepackage{amsmath}
\usepackage{cite}
\usepackage{subfigure}
\usepackage{graphicx}
\usepackage{epstopdf}
\numberwithin{equation}{section}

\usepackage{ctex}
\begin{document}
\title[Rogue wave ]{Rogue waves in a resonant erbium-doped fiber system with higher-order effects}

\author{Yu Zhang, Chuanzhong Li\dag, Jingsong He\ddag$^*$}
\dedicatory {
Department of Mathematics, Ningbo University, Ningbo, 315211 Zhejiang, P.R.China\\
\dag email:lichuanzhong@nbu.edu.cn\\
\ddag email:hejingsong@nbu.edu.cn}
\thanks{$^*$Corresonding Author}
%%%%%%%%%%%%%%%%%%%%%%%%%%%%%%%%%%%%%%%%%%%%%%%%
\begin{abstract}
We mainly investigate a coupled system of the generalized nonlinear Schr\"odinger equation and the Maxwell-Bloch equations which describes the wave propagation in an erbium-doped nonlinear fiber with higher-order effects including the forth-order dispersion and quintic non-Kerr nonlinearity. We derive the one-fold Darbox transformation of this system and construct the determinant representation of the $n$-fold Darboux transformation. Then the determinant representation of the $n$th new solutions $(E^{[n]},\, p^{[n]},\, \eta^{[n]})$ which were generated from the known seed solutions $(E, \, p, \, \eta)$ is established through the $n$-fold Darboux transformation. The solutions $(E^{[n]},\, p^{[n]},\, \eta^{[n]})$ provide the bright and dark breather solutions of this system. Furthermore, we construct the determinant representation of the $n$th-order bright and dark rogue waves by Taylor expansions and also discuss the hybrid solutions which are the nonlinear superposition of the rogue wave and breather solutions.
\end{abstract}
%%%%%%%%%%%%%%%%%%%%%%%%%%%%%%%%%%%%%%%%%%%%%%%%
 \maketitle

PACS numbers: 42.65.Tg, 42.65.Sf, 05.45.Yv, 02.30.Ik.\\
 Keywords: Generalized nonlinear Schr\"odinger and Maxwell-Bloch system; Darboux transformation; Breathers; Rogue waves; Hybrid solutions

%\noindent {\bf PACS} numbers: 02.30.Ik,03.75.Lm,42.65.Tg\\
%02.30.Ik,integrable system;
%03.75.Lm, Tunneling, Josephson effect, Bose¨CEinstein condensates in periodic potentials,
        %solitons, vortices, and topological excitations
%42.65.Tg, Optical solitons; nonlinear guided waves

%%%%%%%%%%%%%%%%%%%%%%%%%%%%%%%%%%%%%%%%%%%%%%%%
\section{Introduction}  \label{sec:1}

Recently, the long haul optical communication has attracted considerable interest of scientists around the world. But the efficiency of propagation of optical communication is still not very well. There are two important reasons, one is because of the dispersion, the other is due to the attenuation. In telecommunications,  what we are interested is the variation of group velocity with frequency, because the absolute wave phase is often not important while the propagation of pulses is important. The dispersion makes the spread of the optical pulse temporally and may lead to the falling of the energy on the next bit slot. The dispersion is the linear effect for the propagation of optical pulse in fibers. The attenuation results from the optical losses which are the inherent feature of the optical fiber. The optical losses also cause the vanishing of the optical pulse due to the absorption and scattering \cite{bib:k.p}.

What is important for the propagation of optical pulses in optical fibers is the nonlinear effect. The optical fiber behaves nonlinearity when the intensity of the optical pulse exceeds a certain threshold value. The most crucial effect is the self-phase modulation (SPM). While traveling in fibers, an optical pulse will induce a varying refractive index of the fiber due to the Kerr effect. This variation in refractive index will produce a phase shift in the pulse which leads to a change of the pulse's frequency spectrum. The spectral broadening process of SPM can balance with the temporal compression due to the anomalous dispersion and reach an equilibrium state when the pulse is of adequate intensity. The resulting pulse is called an optical soliton \cite{bib:SPM}. The possibility of the propagation of optical solitons which are governed by the nonlinear Schr\"odinger (NLS) equation was firstly introduced by Hasegawa and Tappert in 1973 \cite{bib:NLS1,bib:NLS2}. Mollenauer et al. observed experimental solitons in low-loss fiber in 1980 \cite{bib:NLS3}. Another momentous nonlinear effect is the self-induced transparency (SIT). The SIT means that the coherent absorption and re-emission of pulses make the two-level medium optically transparent to this wavelength when the energy difference between the two levels of the medium matches with the optical wavelength. The optical pulse can be amplified and reshaped by passing through an active zone doped with resonant atoms like the erbium. The resonant interaction can neutralize the optical losses in fibers. McCall and Hahn put forward SIT solitons in a two-level resonant system in 1967, which is usually described by the Maxwell-Bloch (MB) equations  \cite{bib:MB1}.

Considering the optical fiber doped with resonant materials such as the erbium, which is governed by a coupled system of the NLS equation and the MB equations,  the  optical pulses satisfy both the  NLS equation and the MB equations. It is necessary to amplify the optical pulses because of the decaying in the process of the propagation in fibers. In 1983 Maimistov and Manykin firstly proposed the NLS-MB system \cite{bib:NLS-MB1}, after that Nakazawa and his cooperators  observed SIT solitons in an erbium-doped silica fiber in 1991 \cite{bib:MB2} and  more studies about the NLS-MB can be seen in Refs. \cite{bib:NLS-MB2,bib:NLS-MB3,bib:NLS-MB4,bib:NLS-MB5}.

Mitschke and Mollenauer found that the observed solitons in experiments did not match with the theoretical properties \cite{bib:OL11}. The propagation equation approximated to a second-order dispersion (group velocity dispersion) with a cubic Kerr nonlinearity (SPM). This leads to the difference between the experimentally observed soliton and NLS soliton in normal fibers. The difference resulted from the additional perturbation of higher order effects such as the higher-order dispersion, self steepening and higher-order nonlinear effects. Therefore the same difference will appear, if one experimentally studies the NLS-MB soliton in erbium-doped optical fibers. So it is necessary to investigate the propagation equation with higher-order effects. The generalized nonlinear Schr\"odinger (GNLS) equation is derived by Porsezian and his partners \cite{bib:GNLS}. The coupled generalized nonlinear Schr\"odinger and Maxwell-Bloch(GNLS-MB) system is described as \cite{bib:GNLS-MB1,bib:GNLS-MB2}
\begin{subequations}\label{eq:GNLSMB}
\begin{align}
E_{z} &= i ( E_{tt}+2 { \left| E \right| }^{2}E)+i\tau (E_{tttt}+8 {\left | E \right |}^2 E_{tt}+2E^2E^{*}_{tt}     \nonumber  \\
& \quad +6E^{*}E^2_{t}+4 {\left | E_{t} \right |}^2E+6 {\left | E \right |}^4 E)+2p , \\
p_{t}& =2i\omega p+2E\eta ,\\
\eta_{t} & = -(Ep^{*}+pE^{*}),
\end{align}
\end{subequations}
where subscripts $z$, $t$ denote as partial derivatives with respect to the distance and time, the asterisk symbol as the complex conjugate, $E$ as the normalized slowly varying amplitude of the complex field envelope, $p=v_{1}v_{2}^{*}$  as the polarization, and $\eta=\left | v_1 \right |^2-\left | v_2 \right |^2$ as the population inversion with $v_1$ and $v_2$ representing the wave functions of the two energy levels of the resonant atoms, $\omega$ as the frequency. The more crucial reason to study the above system in this paper is because of the existence of the quintic non-Kerr nonlinear term which is more significant than the cubic Kerr nonlinearity because the non-Kerr nonlinearity is responsible for the stability of localized solutions \cite{bib:Kerr1,bib:Kerr2}.

In recent years, comparing with solitons, the study on rogue waves in optics has also attracted considerable research due to their potential applications in different branches of physics. The study started from the pioneering measurement of Solli et al. by analyzing super-continuum generations in optical fibers \cite{bib:D.R}. The rogue waves appear from nowhere and disappear without a trace, which is the description of the characteristics of rogue waves \cite{bib:N.A}. The rogue wave occurs for the modulation instability (MI) \cite{bib:MI1,bib:MI2,bib:MI3,bib:MI4,bib:MI5}. One of the possible generating mechanisms for rogue waves is the creation of breathers, then the larger rogue waves can be generated when two or more breathers collide \cite{bib:collide}. The research on rogue waves has made many achievements among which Akhmediev has reported the recent progress in investigating optical rogue waves in Ref. \cite{bib:recent}.

To the best of our knowledge, there are few people to study the GNLS-MB system in the Eq. \eqref{eq:GNLSMB} so far. The solitons and breather solutions of the GNLS-MB system  have been partly established in Ref. \cite{bib:GNLS-MB2}, but the rogue waves of this system is still not reported by anybody. We will construct the determinant representation of the $n$-fold Darboux transformation of the GNLS-MB system, which is similar to the  NLS-MB system \cite{bib:NLS-MB5} and H-MB system \cite{bib:H-MB,bib:2H-MB,bib:IH-MB}. Then the $n$th-order rogue waves of the three optical fields will be given by determinants. Moreover, the $p$ and $\eta$ are found to be dark rogue waves.

The paper is organized as follows. In Section \ref{sec:2}, the Lax pair of the GNLS-MB system is recalled, and we derive the one-fold Darboux transformation of the GNLS-MB system. In Section \ref{sec:3}, the determinant representation of the $n$-fold Darboux transformation and formulas of $(E^{[n]},\, p^{[n]},\, \eta^{[n]})$ are expressed. In Section \ref{sec:4}, the bright and dark breather solutions are generated from  periodic seed solutions. In Section \ref{sec:5}, we construct the determinant representation of the $n$th-order rogue wave by applying Taylor expansions and discuss the effects of parameter $\tau$ on the rogue wave solutions. Additionally, we discuss the hybrid solutions which are the nonlinear superposition of the rogue wave and breather solutions. Finally, we summarize the results in Section \ref{sec:6}.

\section{Lax pair and the one-fold Darboux transformation} \label{sec:2}

In this section, we will derive the one-fold Darboux transformation of the GNLS-MB system in the Eq. \eqref{eq:GNLSMB} of which the Lax pair  is
\begin{subequations}\label{Lax pair}
\begin{align}
\Psi_{t}&=U \Psi, \\
\Psi_{z}&=V \Psi,
\end{align}
\end{subequations}
where%$\Psi=\big(\begin{smallmatrix} \Psi_1 \\ \Psi_2 \end{smallmatrix}\big)$
\begin{subequations}
\begin{align}
\Psi&=\begin{pmatrix} \Psi_1(\lambda;\, t,\, z) \\ \Psi_2(\lambda;\, t,\, z) \end{pmatrix},\\
U&=-i \lambda \sigma_3+U_0,\\
V&=\lambda ^4 V_4+\lambda ^3 V_3+\lambda ^2 V_2+\lambda  V_1+V_0+ \frac{i}{\lambda+\omega}V_{-1},
\end{align}
\end{subequations}
\begin{align*}
\renewcommand\minalignsep{15pt}
\sigma_3&=\begin{pmatrix} 1&0\\0&-1 \end{pmatrix},  \quad U_0=\begin{pmatrix} 0&E\\{-E}^{*} & 0\end{pmatrix} ,\nonumber \\
V_4&=8i\tau  \sigma_3,   \qquad V_3=-8\tau  U_0,\nonumber \\
V_2&=\begin{pmatrix}-2i-4i\tau \left |E \right |^2 & -4i\tau E_t\\-4i\tau E^*_t & 2i+4i\tau \left |E\right| ^2 \end{pmatrix}, \\
V_1&=\begin{pmatrix} 2\tau E^* E_t-2\tau E E^*_t  &  2E+4\tau \left | E \right |^2E+2\tau E_{tt} \\ -2E^*-4\tau \left | E \right |^2E^*-2\tau E^*_{tt}  &-2\tau E^* E_t+
2\tau E E^*_t \end{pmatrix}, \nonumber \\
V_0&=\begin{pmatrix} a_{11} & a_{12}\\a_{21} & -a_{11} \end{pmatrix},   \quad    V_{-1}=\begin{pmatrix} \eta & -p\\-p^* & -\eta \end{pmatrix}, \nonumber \\
\intertext{with}
a_{11}&=i\left | E \right |^2+3i\tau\left | E \right |^4-i\tau \left | E_t \right |^2+i\tau E^* E_{tt}+i\tau E E^*_{tt}, \nonumber \\
a_{12}&=i E_t+6i\tau \left | E \right |^2 E_t+i\tau E_{ttt}, \nonumber \\
a_{21}&=iE^*_t+6i\tau\left | E \right |^2 E^*_t+i\tau E^*_{ttt}.\nonumber
\end{align*}
Here $\lambda$ is a complex  parameter.

Basing on the Darboux transformation for the Ablowitz-Kaup-Newell-Segur(AKNS) system \cite{bib:AKNS}, we consider the following transformation of the Eq. \eqref{eq:GNLSMB},
\begin{align}
\Psi^{[1]}=T_1\Psi=(\lambda I+S)\Psi, \label{DT1}
\intertext{where}
T_1=\lambda I+S,  \nonumber  \qquad  \qquad  \\
I=\begin{pmatrix} 1 & 0  \\  0 & 1 \end{pmatrix},\,
S=\begin{pmatrix} S_{11} & S_{12}  \\  S_{21} & S_{22} \end{pmatrix},\nonumber
\end{align}
such that
\begin{subequations}\label{eq:New1}
\begin{align}
{\Psi^{[1]}}_t=U^{[1]}\Psi^{[1]},\\
{\Psi^{[1]}}_z=V^{[1]}\Psi^{[1]}.
\end{align}
\end{subequations}
Here $U^{[1]},\, V^{[1]}$ depend on $E^{[1]}$, $p^{[1]}$, $\eta^{[1]}$, $\lambda$ and they have the same form as $U$ and $V$ by replacing $E$, $p,\ \eta$ by $E^{[1]}$, $p^{[1]},\, \eta^{[1]}$. In order to make the Eq. \eqref{eq:New1} invariant under the transformation \eqref{DT1}, the $T_1$ must satisfy
\begin{subequations}\label{eq:New2}
\begin{align}
T_{1t}+T_1U=U^{[1]}T_1,\label{eq:New21}\\
T_{1z}+T_1V=V^{[1]}T_1.\label{eq:New22}
\end{align}
\end{subequations}
By comparing the coefficient of $\lambda^i$ $(i=0,\, 1,\, 2)$ on both sides of the Eq. \eqref{eq:New21}, we have
\begin{subequations}
\begin{align}
E^{[1]}&=E+2iS_{12},\quad S_{21}=-S_{12}^*,\\
S_t&=[U_0,\, S]+i[\sigma_3,\, S]S.
\end{align}
\end{subequations}
On the other hand, by multiplying by $(\lambda+\omega)$ and comparing the coefficient of $\lambda^i$ $(i=0,\, 1,\, \dots,\, 6)$ on both sides of the Eq. \eqref{eq:New22}, we get
\begin{subequations}\label{eq:11}
\begin{align}
E^{[1]}&=E+2iS_{12},\\
V^{[1]}_{-1}&=(S-\omega I)V_{-1}(S-\omega I)^{-1} \nonumber \\
&=T_1 |_{\lambda=-\omega} V_{-1}T_1^{-1} |_{\lambda=-\omega}.
\end{align}
\end{subequations}
Additionally, there are some constraints for elements of $V^{[1]}_{-1}$, for instance,  $(V^{[1]}_{-1})_{11}+(V^{[1]}_{-1})_{22} =0$, $\eta^{[1]}= (V^{[1]}_{-1})_{11} \in R$ and $(V^{[1]}_{-1})_{12}^*=(V^{[1]}_{-1})_{21}$. The key step is to find the specific form of S expressed by the column solution of the Eq. \eqref{Lax pair}. Let
\begin{align}\label{eq:12}
S=-H \Lambda H^{-1},
\end{align}
where
\begin{subequations}
\begin{align}
\Lambda &= \begin{pmatrix} \lambda_1 & 0 \\ 0 &\lambda_2 \end{pmatrix},\\
H&=\begin{pmatrix} \Psi_1(\lambda_1;\, t,\, z) & \Psi_1(\lambda_2;\, t,\, z) \\ \Psi_2(\lambda_1;\, t,\, z) & \Psi_2(\lambda_2;\, t,\, z) \end{pmatrix},
\end{align}
\end{subequations}
and $det(H) \neq 0$, $\lambda_1$ and $\lambda_2$ are complex constants.

In order to satisfy the constraints of $S$ and $V^{[1]}_{-1}$ as mentioned above, we take the following constraints
\begin{subequations}\label{eq:yueh}
\begin{align}
\lambda_2&=\lambda_1^*,\\
H&=\begin{pmatrix} \Psi_1(\lambda_1;\, t,\, z) & {-\Psi}^*_2(\lambda_1;\, t,\, z) \\ \Psi_2(\lambda_1;\, t,\, z) & \Psi_1^*(\lambda_1;\, t,\, z) \end{pmatrix}.
\end{align}
\end{subequations}
So after taking the Eqs. \eqref{eq:12} and \eqref{eq:yueh} back into the Eq. \eqref{eq:11}, it results in the following one-fold Darboux transformation of the GNLS-MB system
\begin{subequations}
\begin{align}
E^{[1]}&=E+2iS_{12},\\
p^{[1]}&=-\frac{1}{det(T_1)}[-2\eta (T_1)_{11}(T_1)_{12}+p^*(T_1)_{12}(T_1)_{12}-p(T_1)_{11}(T_1)_{11}] |_{\lambda=-\omega},\\
\eta^{[1]}&=\frac{1}{det(T_1)}[\eta ((T_1)_{11}(T_1)_{22}+(T_1)_{12}(T_1)_{21})-p^*(T_1)_{12}(T_1)_{22}+p(T_1)_{11}(T_1)_{21}]|_{\lambda=-\omega}.
\end{align}
\end{subequations}

\section{ n-fold Darboux transformation for GNLS-MB system} \label{sec:3}

In this section, we will establish the determinant representation of the n-fold Darboux transformation for  the GNLS-MB system as in Ref. \cite{bib:AKNS}. For this purpose, we need to introduce $2n$ eigenfunctions by $f_k=f_k(\lambda_k)=\big(\begin{smallmatrix} f_{k1} \\ f_{k2} \end{smallmatrix} \big)$ associated with an eigenvalue $\lambda_k$, and $\lambda_k \neq \lambda_m$ if $k \neq m$, where $k=1,\, 2,\, \dots,\, 2n$. Additionally, the eigenfunctions for distinct eigenvalues are linearly independent.

According to the form of $T_1$, the n-fold Darboux transformation should have the form $T_n=T_n(\lambda)=\lambda^nI+t_1\lambda^{n-1}+t_2\lambda^{n-2}+\dots+t_{n-1}\lambda+t_n$, where $t_j(j=1,\, 2,\, \dots,\, n)$ are $2\times 2$ matrices. From the following identity
\begin{align}
T_n(\lambda;\, \lambda_1,\, \lambda_2,\, \lambda_3,\, \lambda_4,\, \dots,\, \lambda_{2n-1},\, \lambda_{2n})|_{\lambda=\lambda_k}f_k=0,\, (k=1,\, 2,\, \dots,\, 2n),
\end{align}
we can get the coefficients $t_j(j=1,\, 2,\, \dots,\, n)$ by the Cramer's rule. Thus we obtain the determinant representation of the $T_n$ in the following theorem.

%%%%%%%%%%%%
\newtheorem{theorem}{Theorem}
\begin{theorem}
The n-fold Darboux transformation of the GNLS-MB system is $T_n=T_n(\lambda)=\lambda^nI+t_1\lambda^{n-1}+t_2\lambda^{n-2}+\dots+t_{n-1}\lambda+t_n$, whose determinant representation is
\begin{align} \label{DTn}
T_n=T_n(\lambda;\, \lambda_1,\, \lambda_2,\, \lambda_3,\, \lambda_4,\, \dots,\, \lambda_{2n-1},\, \lambda_{2n})=\frac{1}{|W_{2n}|} \begin{pmatrix} (\widetilde{T_n})_{11} & (\widetilde{T_n})_{12} \\ (\widetilde{T_n})_{21} &(\widetilde{T_n})_{22} \end{pmatrix},
\end{align}
where
\begin{align}
W_{2n}&=\begin{pmatrix}
f_{11} & f_{12} & \lambda_1 f_{11} & \lambda_1 f_{12} & \lambda_1^2 f_{11} & \lambda_1^2 f_{12} & \dots &  \lambda_1^{n-1} f_{11} & \lambda_1^{n-1} f_{12} \\
f_{21} & f_{22} & \lambda_2 f_{21} & \lambda_2 f_{22} & \lambda_2^2 f_{21} & \lambda_2^2 f_{22} & \dots &  \lambda_2^{n-1} f_{21} & \lambda_2^{n-1} f_{22} \\
f_{31} & f_{32} & \lambda_3 f_{31} & \lambda_3 f_{32} & \lambda_3^2 f_{31} & \lambda_3^2 f_{32} & \dots &  \lambda_3^{n-1} f_{31} & \lambda_3^{n-1} f_{32} \\
\vdots & \vdots & \vdots           & \vdots           & \vdots             & \vdots             & \dots &  \vdots                 & \vdots                 \\
f_{2n1} & f_{2n2} & \lambda_{2n} f_{2n1} & \lambda_{2n} f_{2n2} & \lambda_{2n}^2 f_{2n1} & \lambda_{2n}^2 f_{2n2} & \dots &  \lambda_{2n}^{n-1} f_{2n1} & \lambda_{2n}^{n-1} f_{2n2}
\end{pmatrix},  \nonumber \\
\\
(\widetilde{T_n})_{11}&=\begin{vmatrix}
1      & 0      & \lambda          & 0                & \lambda^2          & 0                  & \dots &  \lambda^{n-1}          & 0                      & \lambda^n \\
f_{11} & f_{12} & \lambda_1 f_{11} & \lambda_1 f_{12} & \lambda_1^2 f_{11} & \lambda_1^2 f_{12} & \dots &  \lambda_1^{n-1} f_{11} & \lambda_1^{n-1} f_{12} & \lambda_1^n f_{11} \\
f_{21} & f_{22} & \lambda_2 f_{21} & \lambda_2 f_{22} & \lambda_2^2 f_{21} & \lambda_2^2 f_{22} & \dots &  \lambda_2^{n-1} f_{21} & \lambda_2^{n-1} f_{22} & \lambda_2^n f_{21} \\
f_{31} & f_{32} & \lambda_3 f_{31} & \lambda_3 f_{32} & \lambda_3^2 f_{31} & \lambda_3^2 f_{32} & \dots &  \lambda_3^{n-1} f_{31} & \lambda_3^{n-1} f_{32} & \lambda_3^n f_{31} \\
\vdots & \vdots & \vdots           & \vdots           & \vdots             & \vdots             & \dots &  \vdots                 & \vdots                 & \vdots \\
f_{2n1} & f_{2n2} & \lambda_{2n} f_{2n1} & \lambda_{2n} f_{2n2} & \lambda_{2n}^2 f_{2n1} & \lambda_{2n}^2 f_{2n2} & \dots &  \lambda_{2n}^{n-1} f_{2n1} & \lambda_{2n}^{n-1} f_{2n2} & \lambda_{2n}^n f_{2n1}
\end{vmatrix},   \nonumber  \\
\\
(\widetilde{T_n})_{12}&=\begin{vmatrix}
0      & 1      & 0          & \lambda                & 0          & \lambda^2                  & \dots &  0          & \lambda^{n-1}                      & 0 \\
f_{11} & f_{12} & \lambda_1 f_{11} & \lambda_1 f_{12} & \lambda_1^2 f_{11} & \lambda_1^2 f_{12} & \dots &  \lambda_1^{n-1} f_{11} & \lambda_1^{n-1} f_{12} & \lambda_1^n f_{11} \\
f_{21} & f_{22} & \lambda_2 f_{21} & \lambda_2 f_{22} & \lambda_2^2 f_{21} & \lambda_2^2 f_{22} & \dots &  \lambda_2^{n-1} f_{21} & \lambda_2^{n-1} f_{22} & \lambda_2^n f_{21} \\
f_{31} & f_{32} & \lambda_3 f_{31} & \lambda_3 f_{32} & \lambda_3^2 f_{31} & \lambda_3^2 f_{32} & \dots &  \lambda_3^{n-1} f_{31} & \lambda_3^{n-1} f_{32} & \lambda_3^n f_{31} \\
\vdots & \vdots & \vdots           & \vdots           & \vdots             & \vdots             & \dots &  \vdots                 & \vdots                 & \vdots \\
f_{2n1} & f_{2n2} & \lambda_{2n} f_{2n1} & \lambda_{2n} f_{2n2} & \lambda_{2n}^2 f_{2n1} & \lambda_{2n}^2 f_{2n2} & \dots &  \lambda_{2n}^{n-1} f_{2n1} & \lambda_{2n}^{n-1} f_{2n2} & \lambda_{2n}^n f_{2n1}
\end{vmatrix},   \nonumber  \\
\\
(\widetilde{T_n})_{21}&=\begin{vmatrix}
1      & 0      & \lambda          & 0                & \lambda^2          & 0                  & \dots &  \lambda^{n-1}          & 0                      & 0 \\
f_{11} & f_{12} & \lambda_1 f_{11} & \lambda_1 f_{12} & \lambda_1^2 f_{11} & \lambda_1^2 f_{12} & \dots &  \lambda_1^{n-1} f_{11} & \lambda_1^{n-1} f_{12} & \lambda_1^n f_{12} \\
f_{21} & f_{22} & \lambda_2 f_{21} & \lambda_2 f_{22} & \lambda_2^2 f_{21} & \lambda_2^2 f_{22} & \dots &  \lambda_2^{n-1} f_{21} & \lambda_2^{n-1} f_{22} & \lambda_2^n f_{22} \\
f_{31} & f_{32} & \lambda_3 f_{31} & \lambda_3 f_{32} & \lambda_3^2 f_{31} & \lambda_3^2 f_{32} & \dots &  \lambda_3^{n-1} f_{31} & \lambda_3^{n-1} f_{32} & \lambda_3^n f_{32} \\
\vdots & \vdots & \vdots           & \vdots           & \vdots             & \vdots             & \dots &  \vdots                 & \vdots                 & \vdots \\
f_{2n1} & f_{2n2} & \lambda_{2n} f_{2n1} & \lambda_{2n} f_{2n2} & \lambda_{2n}^2 f_{2n1} & \lambda_{2n}^2 f_{2n2} & \dots &  \lambda_{2n}^{n-1} f_{2n1} & \lambda_{2n}^{n-1} f_{2n2} & \lambda_{2n}^n f_{2n2}
\end{vmatrix},   \nonumber  \\
\\
(\widetilde{T_n})_{22}&=\begin{vmatrix}
0      & 1      & 0          & \lambda                & 0          & \lambda^2                  & \dots &  0          & \lambda^{n-1}                      & \lambda^n \\
f_{11} & f_{12} & \lambda_1 f_{11} & \lambda_1 f_{12} & \lambda_1^2 f_{11} & \lambda_1^2 f_{12} & \dots &  \lambda_1^{n-1} f_{11} & \lambda_1^{n-1} f_{12} & \lambda_1^n f_{12} \\
f_{21} & f_{22} & \lambda_2 f_{21} & \lambda_2 f_{22} & \lambda_2^2 f_{21} & \lambda_2^2 f_{22} & \dots &  \lambda_2^{n-1} f_{21} & \lambda_2^{n-1} f_{22} & \lambda_2^n f_{22} \\
f_{31} & f_{32} & \lambda_3 f_{31} & \lambda_3 f_{32} & \lambda_3^2 f_{31} & \lambda_3^2 f_{32} & \dots &  \lambda_3^{n-1} f_{31} & \lambda_3^{n-1} f_{32} & \lambda_3^n f_{32} \\
\vdots & \vdots & \vdots           & \vdots           & \vdots             & \vdots             & \dots &  \vdots                 & \vdots                 & \vdots \\
f_{2n1} & f_{2n2} & \lambda_{2n} f_{2n1} & \lambda_{2n} f_{2n2} & \lambda_{2n}^2 f_{2n1} & \lambda_{2n}^2 f_{2n2} & \dots &  \lambda_{2n}^{n-1} f_{2n1} & \lambda_{2n}^{n-1} f_{2n2} & \lambda_{2n}^n f_{2n2}
\end{vmatrix}.   \nonumber
\end{align}

\end{theorem}

For the n-fold Darboux transformation, the transformed potentials are
\begin{subequations}
\begin{align}
U_0^{[n]}&=U_0+i[\sigma_3,\, T_n],\\
V_{-1}^{[n]}&=T_n |_{\lambda=-\omega} V_{-1} T_n^{-1} |_{\lambda=-\omega}.
\end{align}
\end{subequations}

If we treat the $n$-fold Darboux transformation as a generalization of the $(n-1)$-fold, it will be multiplications of $n$ one-fold Darboux transformations. It is easy to prove that these multiplications have a determinant representation as mentioned above.

The $n$th new solutions $(E^{[n]},\, p^{[n]},\, \eta^{[n]})$ of the GNLS-MB system after the n-fold Darboux transformation will be
\begin{subequations}\label{nsou}
\begin{align}
E^{[n]}&=E+2i(t_1)_{12}, \\
p^{[n]}&=-\frac{1}{det(T_n)}[-2\eta (T_n)_{11}(T_n)_{12}+p^*(T_n)_{12}(T_n)_{12}-p(T_n)_{11}(T_n)_{11}] |_{\lambda=-\omega},\\
\eta^{[n]}&=\frac{1}{det(T_n)}[\eta ((T_n)_{11}(T_n)_{22}+(T_n)_{12}(T_n)_{21})-p^*(T_n)_{12}(T_n)_{22}+p(T_n)_{11}(T_n)_{21}]|_{\lambda=-\omega}.
\end{align}
\end{subequations}
In order to satisfy the constraints of Darboux transformations, the following conditions must be satisfied
\begin{align} \label{xueshu}
\lambda_{2k}=\lambda_{2k-1}^*, \quad   f_{2k}=\begin{pmatrix} -f_{2k-12}^* \\ f_{2k-11}^* \end{pmatrix}, \quad (k=1,\, 2,\, \dots,\, n).
\end{align}

By calculating, we get $det(T_n)=(\lambda-\lambda_1)(\lambda-\lambda_2)\dots (\lambda-\lambda_{2n-1})(\lambda-\lambda_{2n})$, and $(t_1)_{12}$ in the Eq. \eqref{nsou} is the element of the matrix $t_1$ at the cross of the first row and the second column, where
\begin{align} \label{t1}
t_1&=\frac{1}{|W_{2n}|}\begin{pmatrix} (\widetilde{Q_n})_{11} & (\widetilde{Q_n})_{12} \\ (\widetilde{Q_n})_{21} &(\widetilde{Q_n})_{22} \end{pmatrix},
\end{align}
with
\begin{align}
(\widetilde{Q_n})_{11}&=\begin{vmatrix}
f_{11} & f_{12} & \lambda_1 f_{11} & \lambda_1 f_{12} & \lambda_1^2 f_{11} & \lambda_1^2 f_{12} & \dots &  \lambda_1^{n-1} f_{12} & \lambda_1^{n} f_{11} \\
f_{21} & f_{22} & \lambda_2 f_{21} & \lambda_2 f_{22} & \lambda_2^2 f_{21} & \lambda_2^2 f_{22} & \dots &  \lambda_2^{n-1} f_{22} & \lambda_2^{n} f_{21} \\
f_{31} & f_{32} & \lambda_3 f_{31} & \lambda_3 f_{32} & \lambda_3^2 f_{31} & \lambda_3^2 f_{32} & \dots &  \lambda_3^{n-1} f_{32} & \lambda_3^{n} f_{31} \\
\vdots & \vdots & \vdots           & \vdots           & \vdots             & \vdots             & \dots &  \vdots                 & \vdots                 \\
f_{2n1} & f_{2n2} & \lambda_{2n} f_{2n1} & \lambda_{2n} f_{2n2} & \lambda_{2n}^2 f_{2n1} & \lambda_{2n}^2 f_{2n2} & \dots &  \lambda_{2n}^{n-1} f_{2n2} & \lambda_{2n}^{n} f_{2n1}
\end{vmatrix},  \nonumber \\
\\
(\widetilde{Q_n})_{12}&=-\begin{vmatrix}
f_{11} & f_{12} & \lambda_1 f_{11} & \lambda_1 f_{12} & \lambda_1^2 f_{11} & \lambda_1^2 f_{12} & \dots &  \lambda_1^{n-1} f_{11} & \lambda_1^{n} f_{11} \\
f_{21} & f_{22} & \lambda_2 f_{21} & \lambda_2 f_{22} & \lambda_2^2 f_{21} & \lambda_2^2 f_{22} & \dots &  \lambda_2^{n-1} f_{21} & \lambda_2^{n} f_{21} \\
f_{31} & f_{32} & \lambda_3 f_{31} & \lambda_3 f_{32} & \lambda_3^2 f_{31} & \lambda_3^2 f_{32} & \dots &  \lambda_3^{n-1} f_{31} & \lambda_3^{n} f_{31} \\
\vdots & \vdots & \vdots           & \vdots           & \vdots             & \vdots             & \dots &  \vdots                 & \vdots                 \\
f_{2n1} & f_{2n2} & \lambda_{2n} f_{2n1} & \lambda_{2n} f_{2n2} & \lambda_{2n}^2 f_{2n1} & \lambda_{2n}^2 f_{2n2} & \dots &  \lambda_{2n}^{n-1} f_{2n1} & \lambda_{2n}^{n} f_{2n1}
\end{vmatrix},  \nonumber \\
\\
(\widetilde{Q_n})_{21}&=\begin{vmatrix}
f_{11} & f_{12} & \lambda_1 f_{11} & \lambda_1 f_{12} & \lambda_1^2 f_{11} & \lambda_1^2 f_{12} & \dots &  \lambda_1^{n-1} f_{12} & \lambda_1^{n} f_{12} \\
f_{21} & f_{22} & \lambda_2 f_{21} & \lambda_2 f_{22} & \lambda_2^2 f_{21} & \lambda_2^2 f_{22} & \dots &  \lambda_2^{n-1} f_{22} & \lambda_2^{n} f_{22} \\
f_{31} & f_{32} & \lambda_3 f_{31} & \lambda_3 f_{32} & \lambda_3^2 f_{31} & \lambda_3^2 f_{32} & \dots &  \lambda_3^{n-1} f_{32} & \lambda_3^{n} f_{32} \\
\vdots & \vdots & \vdots           & \vdots           & \vdots             & \vdots             & \dots &  \vdots                 & \vdots                 \\
f_{2n1} & f_{2n2} & \lambda_{2n} f_{2n1} & \lambda_{2n} f_{2n2} & \lambda_{2n}^2 f_{2n1} & \lambda_{2n}^2 f_{2n2} & \dots &  \lambda_{2n}^{n-1} f_{2n2} & \lambda_{2n}^{n} f_{2n2}
\end{vmatrix},  \nonumber \\
\\
(\widetilde{Q_n})_{22}&=-\begin{vmatrix}
f_{11} & f_{12} & \lambda_1 f_{11} & \lambda_1 f_{12} & \lambda_1^2 f_{11} & \lambda_1^2 f_{12} & \dots &  \lambda_1^{n-1} f_{11} & \lambda_1^{n} f_{12} \\
f_{21} & f_{22} & \lambda_2 f_{21} & \lambda_2 f_{22} & \lambda_2^2 f_{21} & \lambda_2^2 f_{22} & \dots &  \lambda_2^{n-1} f_{21} & \lambda_2^{n} f_{22} \\
f_{31} & f_{32} & \lambda_3 f_{31} & \lambda_3 f_{32} & \lambda_3^2 f_{31} & \lambda_3^2 f_{32} & \dots &  \lambda_3^{n-1} f_{31} & \lambda_3^{n} f_{32} \\
\vdots & \vdots & \vdots           & \vdots           & \vdots             & \vdots             & \dots &  \vdots                 & \vdots                 \\
f_{2n1} & f_{2n2} & \lambda_{2n} f_{2n1} & \lambda_{2n} f_{2n2} & \lambda_{2n}^2 f_{2n1} & \lambda_{2n}^2 f_{2n2} & \dots &  \lambda_{2n}^{n-1} f_{2n1} & \lambda_{2n}^{n} f_{2n2}
\end{vmatrix}. \nonumber
\end{align}

So far, we discussed the determinant construction of the $n$th Darboux transformation  of the GNLS-MB system. Then we will constructed breather solutions and rogue wave solutions through these transformations.

\section{bright and dark breather solutions of  GNLS-MB system} \label{sec:4}
In this section, we will focus on breather solutions  $E$, $p$ and $\eta$ of the GNLS-MB system, which are derived from the periodic seed solutions through the Darboux transformation. Then we can get the explicit bright and dark rogue waves of the GNLS-MB system through Taylor expansions of the breather solutions.

\begin{figure}[!htbp]
\centering
\subfigure{\includegraphics[height=4cm,width=5cm]{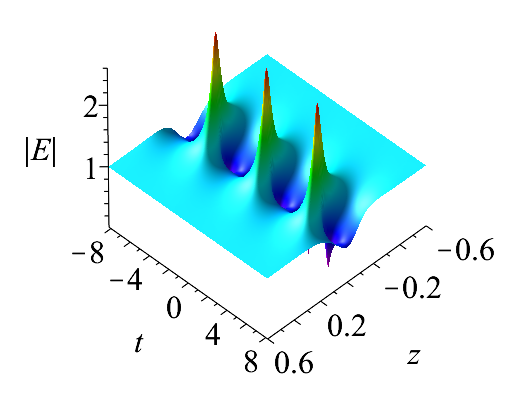}}
\subfigure{\includegraphics[height=4cm,width=5cm]{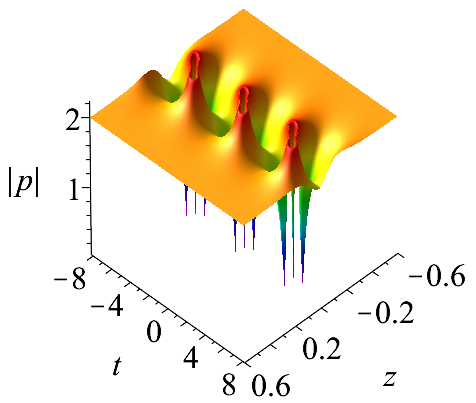}}
\subfigure{\includegraphics[height=4cm,width=5cm]{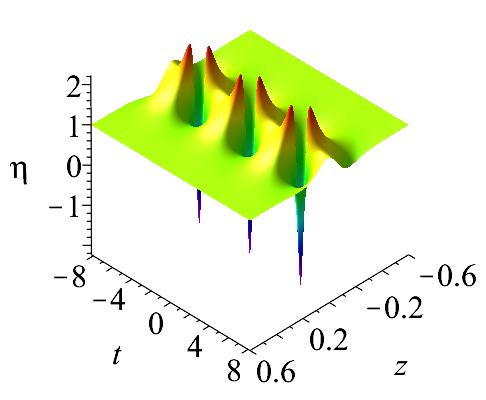}}
\caption{(Color online) The first-order breather solutions $(E,\, p,\, \eta)$ of the GNLS-MB system when $d=1,\, b=2,\, \omega=\frac{1}{2},\, \tau=\frac{1}{2},\, \beta_1=\frac{4}{5}$.}\label{fig:1b}
\end{figure}

Considering the nonzero background wave, we can take $E=de^{i\rho}$, $p=ifE$, and $\eta=1$ as the initial seeds, where $\rho=az+bt$. Then by substituting the seeds into the spectral problem in the Eq. \eqref{Lax pair}, and by separating variables and superposition principle, the eigenfunction $f_{2k-1}$ associated with $\lambda_{2k-1}$ is given by
\begin{align} \label{EF}
f_{2k-1}=\begin{pmatrix} f_{2k-11} \\ f_{2k-12} \end{pmatrix}
        =\begin{pmatrix} C_1 \psi_{2k-11}(\lambda_{2k-1};\, t,\, z)-C_2 \psi_{2k-12}^*(\lambda_{2k-1}^*;\, t,\, z) \\ C_1 \psi_{2k-12}(\lambda_{2k-1};\, t,\, z)+C_2 \psi_{2k-11}^*(\lambda_{2k-1}^*;\, t,\, z) \end{pmatrix},
\end{align}
where
\begin{align}
\psi_{2k-1}(\lambda_{2k-1};\, t,\, z)=\begin{pmatrix} \psi_{2k-11}(\lambda_{2k-1};\, t,\, z) \\ \psi_{2k-12}(\lambda_{2k-1};\, t,\, z) \end{pmatrix}
=\begin{pmatrix} de^{(\frac{i}{2} \rho+ic(\lambda_{2k-1}))} \\ i(c_1(\lambda_{2k-1})+\lambda_{2k-1}+\frac{b}{2})e^{-\frac{i}{2}\rho+ic(\lambda_{2k-1})} \end{pmatrix}. \nonumber
\end{align}
Here $\psi_{2k-1}(\lambda_{2k-1};\, t,\, z)$ is the basic solution of the spectral problem in the Eq. \eqref{Lax pair}, with $a,\, b,\, d,\, z,\, t \in \mathbb{R}$, $C_1,\, C_2 \in \mathbb{C}$
\begin{align}
a=&\tau (b^4+6d^4-12b^2d^2)-b^2+2d^2+2f, \nonumber  \\
f=&\frac{2}{2\omega-b}, \nonumber \\
c(\lambda_{2k-1})=&c_1(\lambda_{2k-1})z+c_2(\lambda_{2k-1})t, \nonumber  \\
c_1(\lambda_{2k-1})=&\{(2\lambda_{2k-1}-b)+\tau [ -8\lambda_{2k-1}^3+4b\lambda_{2k-1}^2+(4d^2-2b^2)\lambda_{2k-1}   \nonumber \\
&-6d^2b+b^3]+\frac{f}{ \lambda_{2k-1}+\omega} \}c_2(\lambda_{2k-1}), \nonumber \\
c_2(\lambda_{2k-1})=&\sqrt{d^2+(\lambda_{2k-1}+\frac{b}{2})^2}. \nonumber
\end{align}
Without lose of generality, in above identities we consider the case when $C_1=C_2=1$, $\lambda_{2k-1}=-\frac{b}{2}+i\beta_{2k-1}$, such that $c_2(\lambda_{2k-1})^2=d^2-Im^2(\lambda_{2k-1}) \in \mathbb{R}$. Under the condition in the Eq. \eqref{xueshu}, by substituting eigenfunctions in the Eq. \eqref{EF} into the Eq. \eqref{nsou} with the choice of the Eq. \eqref{DTn} and the Eq. \eqref{t1}, we can construct the $n$th-order solutions $E^{[n]},\, p^{[n]}$ and $\eta^{[n]}$.  When $n=1$, here we just give an expression of the breather solution $E^{[1]}_b$ with the specific parameters $d=1,\, b=2,\, \omega=\frac{1}{2},\, \tau=\frac{1}{2}$ in the following,
\begin{align}
E^{[1]}_b=-{\frac {\beta_1 \cos (\theta_1)+(1-2\beta_1^2) \cosh (\theta_2)+2i \beta_1 K_0 \sinh (\theta_2)}{\beta_{{1}}\cos \left( {\theta_1} \right) -\cosh \left( {\theta_2} \right) }}e^{i(-19z+2t)},
\end{align}
where
\begin{align}
K_0&=\sqrt{1-\beta_1^2}, \nonumber \\
\theta_1&=\frac{2K_0[(-64\beta_1^4+8)z+(4\beta_1^2+1)t]}{4\beta_1^2+1}, \nonumber \\
\theta_2&=\frac{8 \beta_1 K_0  (4 \beta_1^4-19 \beta_1^2-3)z}{4\beta_1^2+1}. \nonumber
\end{align}
Note that the trajectory of $E^{[1]}_b$ is defined by
\begin{align}
(-64\beta_1^4+8)z+(4\beta_1^2+1)t=0, \nonumber
\end{align}
if $K_0^2<0$, and by
\begin{align}
\beta_1 (4 \beta_1^4-19 \beta_1^2-3)z=0, \nonumber
\end{align}
if $K_0^2>0$. When $\beta_1=\frac{4}{5}$, we have $K_0^2>0$ then can get the temporal periodic breather solution (Ma breather \cite{bib:Mb}). Meanwhile the breather solutions $p^{[1]}_b$ and $\eta^{[1]}_b$ can be constructed with the same specific parameters of $E^{[1]}_b$. The dynamical evolution of the Ma breather solutions is plotted in Fig. \ref{fig:1b}.

After a simple analysis, we can know from the eigenfunction that $\tau$ has an influence on the periodic of the breather solutions except Ma breather solutions. Having constructed a bright breather for $E$ and dark breathers for $p$ and $\eta$ when $K_0^2 \neq 0$, in the next section, our main object is to discuss the construction of the rogue wave solutions of the GNLS-MB system when $K_0^2$ goes to zero.

\section{bright and dark rogue waves of  GNLS-MB system} \label{sec:5}

In this section, firstly, we will construct the first-order bright and dark rogue waves of the GNLS-MB system by using the limit method. This kind of solutions only appears in some special regions of distance and time and then will be drowned in one fixed non-vanishing plane. Under the condition $C_1=C_2=1$, substituting eigenfunctions in the Eq. \eqref{EF} into the Eq. \eqref{nsou} with $\lambda_1=-\frac{b}{2}+i\beta_1$, by taking the limit $\beta_1 \rightarrow d \,(d>0)$, $E^{[1]},\, p^{[1]}$ and $\eta^{[1]}$ become rational solutions  $E^{[1]}_r,\, p^{[1]}_r$ and $\eta^{[1]}_r$ in the form of rogue waves \cite{bib:NLS-MB4}. When $z\rightarrow \infty,\, t \rightarrow \infty$ in the expressions of rogue waves, by calculation, we find that the non-vanishing  background plane of $E^{[1]}_r,\,  p^{[1]}_r$ and $\eta^{[1]}_r$ has nothing to do with $\tau$, i.e.
\begin{align}
|E^{[1]}_r| \rightarrow d, \qquad  |p^{[1]}_r| \rightarrow \frac{2d}{|b-2\omega|}, \qquad  \eta^{[1]}_r \rightarrow 1.
\end{align}

Because the solutions are very complicated, we take $d=1,\, b=2,\, \omega=\frac{1}{2}$ in order to display  the results easily. Then the final forms of the rogue waves will be
\begin{subequations} \label{1r}
\begin{align}
E^{[1]}_r&=e^{2i[-(13\tau+3)z+t]}(-1+\frac{k_2}{k_1}),     \label{1rE}           \\
p^{[1]}_r&=2ie^{2i[-(13\tau+3)z+t]}(1+\frac{4k_3+16k_4i}{k_1^2}),   \label{1rp}   \\
\eta^{[1]}_r&=1+\frac{16k_5}{k_1^2},  \label{1reta}
\end{align}
\end{subequations}
where
\begin{align*}
k_1&=16s_1z^2-128s_2zt+20t^2+5, \\
 k_2&=20-144s_3iz,         \\
k_3&=-16(17780\tau^2-2500\tau+101)z^2+160s_4zt-20t^2-5,           \\
k_4&=144s_1s_3z^3-1152s_2s_3z^2t+180s_3zt^2-5(86\tau+1)z+10t,     \\
k_5&=16(1580\tau^2+740\tau-61)z^2-160s_4zt+20t^2-5,
\end{align*}
with
\begin{align}\label{s}
s_1=1940\tau^2-196\tau+29,   \quad     s_2=5\tau+1,   \quad   s_3=10\tau-1,  \quad    s_4=22\tau-1.
\end{align}

In the Eq. \eqref{1rE}, from $|E^{[1]}_r|_z=0$, and $|E^{[1]}_r|_t=0$, we get that the maximum amplitude of $|E^{[1]}_r|$ occurs at $z=0,\, t=0$ and is equal to $3$. The minimum occurs at $z=0,\, t=\pm \frac{\sqrt{3}}{2}$ and is equal to $0$. We also infer that $|E^{[1]}_r| \rightarrow 1$ by assuming $z\rightarrow \infty$, $t\rightarrow \infty$, which gives the background plane.

In the Eq. \eqref{1rp}, by the same method we get that the height of the background plane is $2$, for $|p^{[1]}_r| \rightarrow 2$, when $z\rightarrow \infty$,\, $t\rightarrow \infty$. The maximum amplitude of $|p^{[1]}_r|$ occurs in the form of upper ring curve  as
\begin{align}
256s_1z^4-4096s_1s_2z^3t+384(4300\tau^2+100\tau+91)z^2t^2-5120s_2zt^3+400t^4      \nonumber \\
  +32(22340\tau^2+4940\tau-343)z^2-1280(49\tau-1)zt+520t^2-55=0,
\end{align}
the maximum amplitude is equal to $\sqrt{5}$. The minimum amplitude of $|p^{[1]}_{r}|$ occurs in terms of four down peaks,
\begin{align}
\big(z_1&= \quad   \frac{5  -   \sqrt{5}}{72(10\tau-1)},  \,   t_1=  \quad   \frac{(590\tau-71)   -  (58\tau-37)\sqrt{5}}{36(10\tau-1)(3\sqrt{5}   +   2)}\big), \nonumber \\
\big(z_2&= \quad   \frac{5  +   \sqrt{5}}{72(10\tau-1)},  \,   t_2=   -      \frac{(590\tau-71)   +  (58\tau-37)\sqrt{5}}{36(10\tau-1)(3\sqrt{5}   -   2)}\big), \nonumber \\
\big(z_3&= -       \frac{5  -   \sqrt{5}}{72(10\tau-1)},  \,   t_3=   -      \frac{(590\tau-71)   -  (58\tau-37)\sqrt{5}}{36(10\tau-1)(3\sqrt{5}   +   2)}\big), \nonumber \\
\big(z_4&= -       \frac{5  +   \sqrt{5}}{72(10\tau-1)},  \,   t_4=  \quad   \frac{(590\tau-71)   +  (58\tau-37)\sqrt{5}}{36(10\tau-1)(3\sqrt{5}   -   2)}\big),\nonumber
\end{align}
and is equal to $0$. Meanwhile the extreme of the amplitude $|p^{[1]}_r|$ occurs at $z=0,\, t=0$ and is equal to $\frac{2}{5}$.

In the Eq. \eqref{1reta}, we conclude that the height of the background plane is $1$ because $\eta^{[1]}_r \rightarrow 1$ when $z\rightarrow \infty$,\, $t\rightarrow \infty$. The  $\eta^{[1]}_r$ has  two upper peaks at coordinates as
\begin{align}
\big(z_2&= \quad   \frac{5  +   \sqrt{5}}{72(10\tau-1)},  \,  t_2=   -      \frac{(590\tau-71)   +  (58\tau-37)\sqrt{5}}{36(10\tau-1)(3\sqrt{5}   -   2)}\big), \nonumber \\
\big(z_4&= -       \frac{5  +   \sqrt{5}}{72(10\tau-1)},  \,   t_4=  \quad   \frac{(590\tau-71)   +  (58\tau-37)\sqrt{5}}{36(10\tau-1)(3\sqrt{5}   -   2)}\big), \nonumber
\end{align}
and the height is equal to $\sqrt{5}$. The minimum amplitude of $\eta^{[1]}_r$ occurs at two down peaks with coordinates as
\begin{align}
\big(z_1&= \quad   \frac{5  -   \sqrt{5}}{72(10\tau-1)},  \,   t_1=  \quad   \frac{(590\tau-71)   -  (58\tau-37)\sqrt{5}}{36(10\tau-1)(3\sqrt{5}   +   2)}\big), \nonumber \\
\big(z_3&= -       \frac{5  -   \sqrt{5}}{72(10\tau-1)},  \,   t_3=   -      \frac{(590\tau-71)   -  (58\tau-37)\sqrt{5}}{36(10\tau-1)(3\sqrt{5}   +   2)}\big), \nonumber
\end{align}
and is equal to $-\sqrt{5}$. There is another extreme of the amplitude $\eta^{[1]}_r$ which occurs at $z=0,\, t=0$ and is equal to $-\frac{11}{5}$.

\begin{figure}[!htbp]
\centering
\subfigure{\includegraphics[height=4cm,width=5cm]{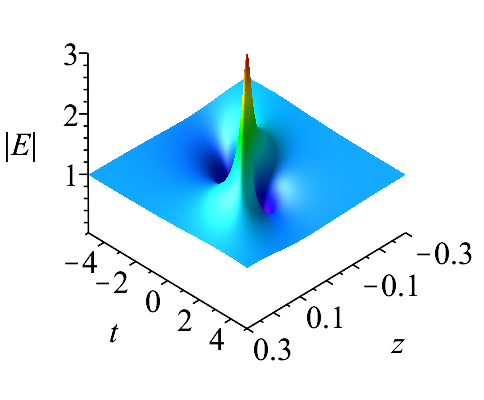}}
\subfigure{\includegraphics[height=4cm,width=5cm]{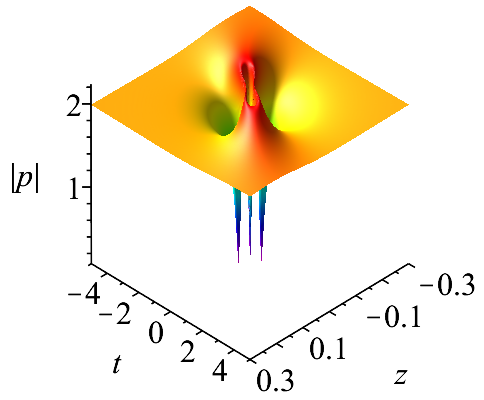}}
\subfigure{\includegraphics[height=4cm,width=5cm]{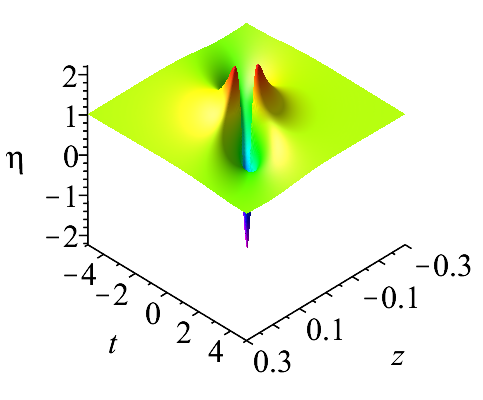}}
\caption{(Color online) The first-order rogue wave solutions $(E,\, p,\, \eta)$ of the GNLS-MB system when $d=1,\, b=2,\, \omega=\frac{1}{2},\, \tau=\frac{1}{2}$.}\label{fig:1r}
\end{figure}

The first-order rogue wave solutions in the Eq. \eqref{1r} are plotted in Fig. \ref{fig:1r} from which we just find three down peaks in the second sub-figure of the Fig. \ref{fig:1r}. On account of the direction of the observation, the two close sub-peaks of the middle down peak are not distinguished clearly from the sub-figure. For the same reason, we only catch sight of one down peak in the third sub-figure in Fig. \ref{fig:1r}.

\begin{figure}[!htbp]
\centering
\raisebox{20 ex}{(a)}\subfigure{\includegraphics[height=4cm,width=5cm]{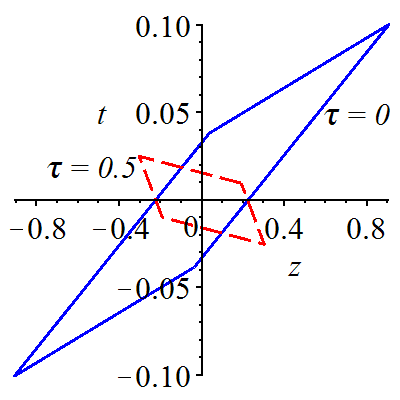}}
\raisebox{20 ex}{(b)}\subfigure{\includegraphics[height=4cm,width=5cm]{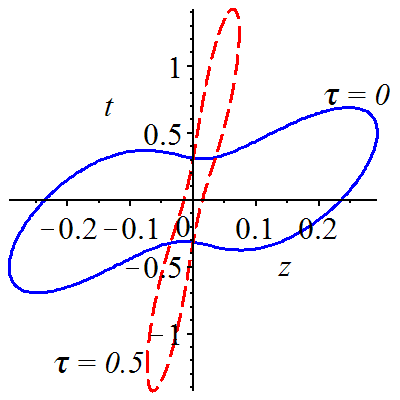}}
\raisebox{20 ex}{(c)}\subfigure{\includegraphics[height=4cm,width=5cm]{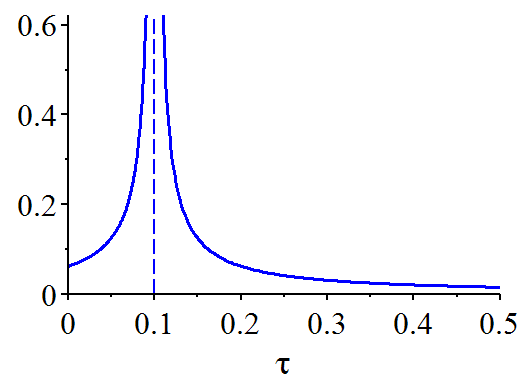}}
\caption{(Color online) When $d=1,\, b=2,\, \omega=\frac{1}{2}$: $(a)$ the parallelogram consists of the four extreme points $(z_i,\, t_i)$ with $\tau=0$ (blue, solid) and $\tau=\frac{1}{2}$ (red, dash). $(b)$ the upper ring curve of the $|p^{[1]}_r|$ with $\tau=0$ (blue, solid) and $\tau=\frac{1}{2}$ (red, dash). $(c)$ the area function of the parallelogram in $(a)$ with different $\tau$.}\label{fig:p1}
\end{figure}

By the calculation above, we find that $\tau$ has an impact on the location of extreme  except when $z=0$, but does not change the value of the extreme. We can get a parallelogram by connecting the four extreme points $(z_i,\, t_i)$ with $i=1,\, 2,\, 3,\, 4$ sequentially. The change of the parallelogram reflects the compression and rotation of the rogue wave. The parallelogram is plotted in Fig. \ref{fig:p1}$(a)$ with different $\tau$. In order to reflect the change of the rogue wave more clearly, we also give the upper ring curve of $|p^{[1]}_r|$ in Fig. \ref{fig:p1}$(b)$ with the same $\tau$ in Fig. \ref{fig:p1}$(a)$. The area of the parallelogram is equal to $\frac{\sqrt{5}}{36|10\tau-1|}$ and its function is plotted in Fig. \ref{fig:p1}$(c)$.

In order to enrich the type of the $n$th-order rogues, we can modify $C_1$ and $C_2$ in the Eq. \eqref{EF} as following,
\begin{subequations}\label{rwC}
\begin{align}
C_1&=e^{ic_2(\lambda_{2k-1}) \sum_{j=0}^{n-1}J_j[\lambda_{2k-1}-(-\frac{b}{2}+id)]^j}, \\
C_2&=e^{-ic_2(\lambda_{2k-1}) \sum_{j=0}^{n-1}J_j[\lambda_{2k-1}-(-\frac{b}{2}+id)]^j},
\end{align}
\end{subequations}
where $J_j \in \mathbb{C}$. Note that $\lambda_{2k-1}=-\frac{b}{2}+id$ is the zero point of $c_2(\lambda_{2k-1})$.

We can get $n$th-order rogue waves by letting $\lambda_{2k-1} \rightarrow -\frac{b}{2}+id$ in the $n$th-order breather solutions. But it is difficult to obtain higher-order rogue waves from multi-breather solutions generally, because the explicit expression of the higher-order breather is very difficult to calculate. In order to solve the problem, we derive the $n$th-order rogue waves directly from the determinant representations of  solutions in the Eq. \eqref{nsou} by applying Taylor expansions \cite{bib:limit}.

\begin{theorem}
For the $n$-fold Darboux transformation, substituting the Eq. \eqref{rwC} into the eigenfunction in the Eq. \eqref{EF}, letting $\lambda_{2k-1}=-\frac{b}{2}+id+\varepsilon^2 (\varepsilon>0)$, by adopting Taylor expansions, the determinant representations of the $n$th-order rogue waves of the GNLS-MB system are given as
\begin{subequations}\label{nrw}
\begin{align}
E^{[n]}_r&=E+2i(t_{r1})_{12}, \\
p^{[n]}_r&=-\frac{1}{det(T_{rn})}[-2\eta (T_{rn})_{11}(T_{rn})_{12}+p^*(T_{rn})_{12}(T_{rn})_{12}-p(T_{rn})_{11}(T_{rn})_{11}] |_{\lambda=-\omega},\\
\eta^{[n]}_r&=\frac{1}{det(T_{rn})}[\eta ((T_{rn})_{11}(T_{rn})_{22}+(T_{rn})_{12}(T_{rn})_{21})-p^*(T_{rn})_{12}(T_{rn})_{22}+p(T_{rn})_{11}(T_{rn})_{21}]|_{\lambda=-\omega},
\end{align}
\end{subequations}
where
\begin{align}
T_{rn}=\frac{1}{|W_{r2n}|} \begin{pmatrix} (\widetilde{T_{rn}})_{11} & (\widetilde{T_{rn}})_{12} \\ (\widetilde{T_{rn}})_{21} &(\widetilde{T_{rn}})_{22} \end{pmatrix}, \,
t_{r1}=\frac{1}{|W_{r2n}|}\begin{pmatrix} (\widetilde{Q_{rn}})_{11} & (\widetilde{Q_{rn}})_{12} \\ (\widetilde{Q_{rn}})_{21} &(\widetilde{Q_{rn}})_{22} \end{pmatrix},
\end{align}
with
\begin{align*}
W_{r2n}&=\begin{pmatrix}
h^1_{01}         &  h^1_{02}        &  h^1_{11}        &  h^1_{12}         &  \cdots   &   h^1_{n-11}        &  h^1_{n-12}          \\
-h^{1*}_{02}     &  h^{1*}_{01}     &  -h^{1*}_{12}    &  h^{1*}_{11}      &  \cdots   &   -h^{1*}_{n-12}    &  h^{1*}_{n-11}       \\
h^3_{01}         &  h^3_{02}        &  h^3_{11}        &  h^3_{12}         &  \cdots   &   h^3_{n-11}        &  h^3_{n-12}          \\
\vdots           &  \vdots          &  \vdots          &  \vdots           &  \vdots   &   \vdots            &  \vdots              \\
-h^{2n-1*}_{02}  &  h^{2n-1*}_{01}  &  -h^{2n-1*}_{12} &  h^{2n-1*}_{11}   &  \cdots   &   -h^{2n-1*}_{n-12} &  h^{2n-1*}_{n-11}
\end{pmatrix},\\
\\
(\widetilde{T_{rn}})_{11}&=\begin{vmatrix}
1                &  0               &  \lambda         &  0                &  \cdots   &   \lambda^{n-1}     &  0                   &   \lambda^{n}     \\
h^1_{01}         &  h^1_{02}        &  h^1_{11}        &  h^1_{12}         &  \cdots   &   h^1_{n-11}        &  h^1_{n-12}          &   h^1_{n1}        \\
-h^{1*}_{02}     &  h^{1*}_{01}     &  -h^{1*}_{12}    &  h^{1*}_{11}      &  \cdots   &   -h^{1*}_{n-12}    &  h^{1*}_{n-11}       &   -h^{1*}_{n2}    \\
h^3_{01}         &  h^3_{02}        &  h^3_{11}        &  h^3_{12}         &  \cdots   &   h^3_{n-11}        &  h^3_{n-12}          &   h^3_{n1}        \\
\vdots           &  \vdots          &  \vdots          &  \vdots           &  \vdots   &   \vdots            &  \vdots              &   \vdots          \\
-h^{2n-1*}_{02}  &  h^{2n-1*}_{01}  &  -h^{2n-1*}_{12} &  h^{2n-1*}_{11}   &  \cdots   &   -h^{2n-1*}_{n-12} &  h^{2n-1*}_{n-11}    &   -h^{2n-1*}_{n2}
\end{vmatrix},\\
\\
(\widetilde{T_{rn}})_{12}&=\begin{vmatrix}
0                &  1               &  0               &  \lambda          &  \cdots   &   0                 &  \lambda^{n-1}       &   0               \\
h^1_{01}         &  h^1_{02}        &  h^1_{11}        &  h^1_{12}         &  \cdots   &   h^1_{n-11}        &  h^1_{n-12}          &   h^1_{n1}        \\
-h^{1*}_{02}     &  h^{1*}_{01}     &  -h^{1*}_{12}    &  h^{1*}_{11}      &  \cdots   &   -h^{1*}_{n-12}    &  h^{1*}_{n-11}       &   -h^{1*}_{n2}    \\
h^3_{01}         &  h^3_{02}        &  h^3_{11}        &  h^3_{12}         &  \cdots   &   h^3_{n-11}        &  h^3_{n-12}          &   h^3_{n1}        \\
\vdots           &  \vdots          &  \vdots          &  \vdots           &  \vdots   &   \vdots            &  \vdots              &   \vdots          \\
-h^{2n-1*}_{02}  &  h^{2n-1*}_{01}  &  -h^{2n-1*}_{12} &  h^{2n-1*}_{11}   &  \cdots   &   -h^{2n-1*}_{n-12} &  h^{2n-1*}_{n-11}    &   -h^{2n-1*}_{n2}
\end{vmatrix},\\
\\
(\widetilde{T_{rn}})_{21}&=\begin{vmatrix}
1                &  0               &  \lambda         &  0                &  \cdots   &   \lambda^{n-1}     &  0                   &   0               \\
h^1_{01}         &  h^1_{02}        &  h^1_{11}        &  h^1_{12}         &  \cdots   &   h^1_{n-11}        &  h^1_{n-12}          &   h^1_{n2}        \\
-h^{1*}_{02}     &  h^{1*}_{01}     &  -h^{1*}_{12}    &  h^{1*}_{11}      &  \cdots   &   -h^{1*}_{n-12}    &  h^{1*}_{n-11}       &   h^{1*}_{n1}     \\
h^3_{01}         &  h^3_{02}        &  h^3_{11}        &  h^3_{12}         &  \cdots   &   h^3_{n-11}        &  h^3_{n-12}          &   h^3_{n2}        \\
\vdots           &  \vdots          &  \vdots          &  \vdots           &  \vdots   &   \vdots            &  \vdots              &   \vdots          \\
-h^{2n-1*}_{02}  &  h^{2n-1*}_{01}  &  -h^{2n-1*}_{12} &  h^{2n-1*}_{11}   &  \cdots   &   -h^{2n-1*}_{n-12} &  h^{2n-1*}_{n-11}    &   h^{2n-1*}_{n1}
\end{vmatrix},\\
\\
(\widetilde{T_{rn}})_{22}&=\begin{vmatrix}
0                &  1               &  0               &  \lambda          &  \cdots   &   0                 &  \lambda^{n-1}       &   \lambda^n       \\
h^1_{01}         &  h^1_{02}        &  h^1_{11}        &  h^1_{12}         &  \cdots   &   h^1_{n-11}        &  h^1_{n-12}          &   h^1_{n2}        \\
-h^{1*}_{02}     &  h^{1*}_{01}     &  -h^{1*}_{12}    &  h^{1*}_{11}      &  \cdots   &   -h^{1*}_{n-12}    &  h^{1*}_{n-11}       &   h^{1*}_{n1}     \\
h^3_{01}         &  h^3_{02}        &  h^3_{11}        &  h^3_{12}         &  \cdots   &   h^3_{n-11}        &  h^3_{n-12}          &   h^3_{n2}        \\
\vdots           &  \vdots          &  \vdots          &  \vdots           &  \vdots   &   \vdots            &  \vdots              &   \vdots          \\
-h^{2n-1*}_{02}  &  h^{2n-1*}_{01}  &  -h^{2n-1*}_{12} &  h^{2n-1*}_{11}   &  \cdots   &   -h^{2n-1*}_{n-12} &  h^{2n-1*}_{n-11}    &   h^{2n-1*}_{n1}
\end{vmatrix},\\
\\
(\widetilde{Q_{rn}})_{11}&=\begin{vmatrix}
h^1_{01}         &  h^1_{02}        &  h^1_{11}        &  h^1_{12}         &  \cdots   &  h^1_{n-12}         &   h^1_{n1}          \\
-h^{1*}_{02}     &  h^{1*}_{01}     &  -h^{1*}_{12}    &  h^{1*}_{11}      &  \cdots   &  h^{1*}_{n-11}      &   -h^{1*}_{n2}      \\
h^3_{01}         &  h^3_{02}        &  h^3_{11}        &  h^3_{12}         &  \cdots   &  h^3_{n-12}         &   h^3_{n1}          \\
\vdots           &  \vdots          &  \vdots          &  \vdots           &  \vdots   &  \vdots             &   \vdots            \\
-h^{2n-1*}_{02}  &  h^{2n-1*}_{01}  &  -h^{2n-1*}_{12} &  h^{2n-1*}_{11}   &  \cdots   &  h^{2n-1*}_{n-11}   &   -h^{2n-1*}_{n2}
\end{vmatrix},\\
\\
(\widetilde{Q_{rn}})_{12}&=-\begin{vmatrix}
h^1_{01}         &  h^1_{02}        &  h^1_{11}        &  h^1_{12}         &  \cdots   &  h^1_{n-11}         &   h^1_{n1}          \\
-h^{1*}_{02}     &  h^{1*}_{01}     &  -h^{1*}_{12}    &  h^{1*}_{11}      &  \cdots   &  -h^{1*}_{n-12}     &   -h^{1*}_{n2}      \\
h^3_{01}         &  h^3_{02}        &  h^3_{11}        &  h^3_{12}         &  \cdots   &  h^3_{n-11}         &   h^3_{n1}          \\
\vdots           &  \vdots          &  \vdots          &  \vdots           &  \vdots   &  \vdots             &   \vdots            \\
-h^{2n-1*}_{02}  &  h^{2n-1*}_{01}  &  -h^{2n-1*}_{12} &  h^{2n-1*}_{11}   &  \cdots   &  -h^{2n-1*}_{n-12}  &   -h^{2n-1*}_{n2}
\end{vmatrix},\\
\\
(\widetilde{Q_{rn}})_{21}&=\begin{vmatrix}
h^1_{01}         &  h^1_{02}        &  h^1_{11}        &  h^1_{12}         &  \cdots   &  h^1_{n-12}         &   h^1_{n2}          \\
-h^{1*}_{02}     &  h^{1*}_{01}     &  -h^{1*}_{12}    &  h^{1*}_{11}      &  \cdots   &  h^{1*}_{n-11}      &   h^{1*}_{n1}       \\
h^3_{01}         &  h^3_{02}        &  h^3_{11}        &  h^3_{12}         &  \cdots   &  h^3_{n-12}         &   h^3_{n2}          \\
\vdots           &  \vdots          &  \vdots          &  \vdots           &  \vdots   &  \vdots             &   \vdots            \\
-h^{2n-1*}_{02}  &  h^{2n-1*}_{01}  &  -h^{2n-1*}_{12} &  h^{2n-1*}_{11}   &  \cdots   &  h^{2n-1*}_{n-11}   &   h^{2n-1*}_{n1}
\end{vmatrix},\\
\\
(\widetilde{Q_{rn}})_{22}&=-\begin{vmatrix}
h^1_{01}         &  h^1_{02}        &  h^1_{11}        &  h^1_{12}         &  \cdots   &  h^1_{n-11}         &   h^1_{n2}          \\
-h^{1*}_{02}     &  h^{1*}_{01}     &  -h^{1*}_{12}    &  h^{1*}_{11}      &  \cdots   &  -h^{1*}_{n-12}     &   h^{1*}_{n1}      \\
h^3_{01}         &  h^3_{02}        &  h^3_{11}        &  h^3_{12}         &  \cdots   &  h^3_{n-11}         &   h^3_{n2}          \\
\vdots           &  \vdots          &  \vdots          &  \vdots           &  \vdots   &  \vdots             &   \vdots            \\
-h^{2n-1*}_{02}  &  h^{2n-1*}_{01}  &  -h^{2n-1*}_{12} &  h^{2n-1*}_{11}   &  \cdots   &  -h^{2n-1*}_{n-12}  &   h^{2n-1*}_{n1}
\end{vmatrix},
\end{align*}
\begin{align*}
h^l_{mj}=\frac{\partial^l}{\partial \varepsilon ^l} \big [(-\frac{b}{2}+id+\varepsilon^2)^m f_{1j}(\lambda_1=-\frac{b}{2}+id+\varepsilon^2) \big] \Big |_{\varepsilon=0}, \\
  \qquad  m=0,\, 1,\, 2,\, \cdots,\, n; \quad j=1,\, 2;  \quad  l=1,\, 2,\, \cdots,\, 2n.
\end{align*}

\end{theorem}

According to our analysis, we can know that there are $n+4$ free parameters denoted as $(J_0,\, J_1,\, \cdots,\, J_{n-1};\, b,\, d,\, \omega,\, \tau)$ in the $n$th-order rogue wave solutions. Next, we will consider the types of the $n$th-order rogue waves with these parameters. For convenience, we let $d=1,\, b=0,\, \omega=\frac{1}{2},\, \tau=\frac{1}{2}$ in the following.

\begin{figure}[!htbp]
\centering
\subfigure{\includegraphics[height=4cm,width=5cm]{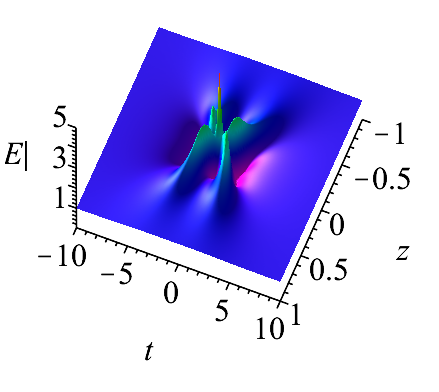}}
\subfigure{\includegraphics[height=4cm,width=5cm]{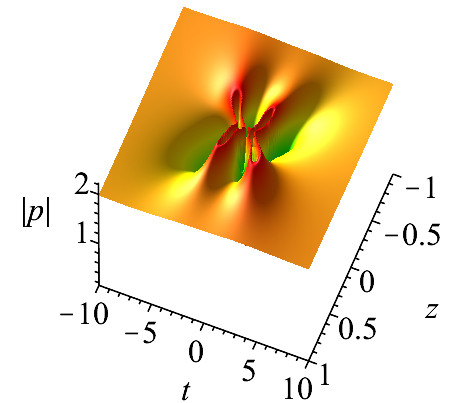}}
\subfigure{\includegraphics[height=4cm,width=5cm]{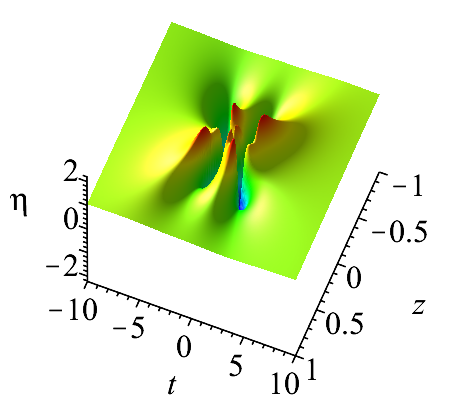}}
\caption{(Color online) The fundamental pattern of the second-order rogue wave solutions $(E,p,\eta)$ of the GNLS-MB system when $d=1, b=0, \omega=\frac{1}{2}, \tau=\frac{1}{2}, J_0=0, J_1=0$.}\label{fig:2r1}
\end{figure}

\begin{figure}[!htbp]
\centering
\subfigure{\includegraphics[height=4cm,width=5cm]{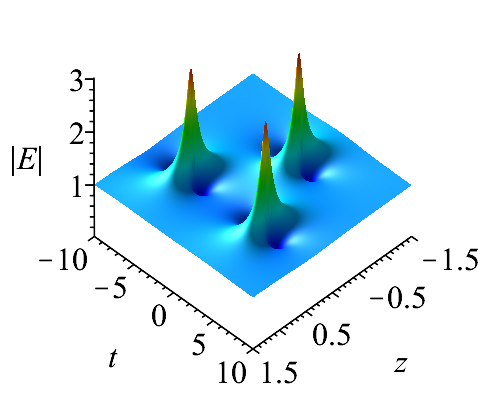}}
\subfigure{\includegraphics[height=4cm,width=5cm]{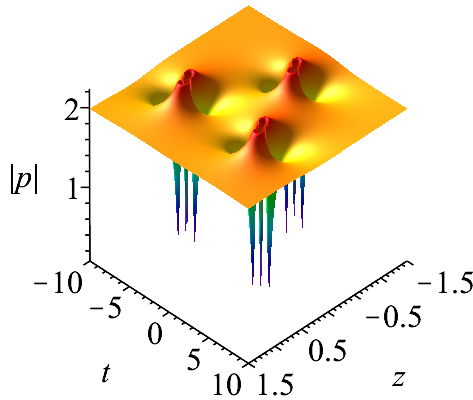}}
\subfigure{\includegraphics[height=4cm,width=5cm]{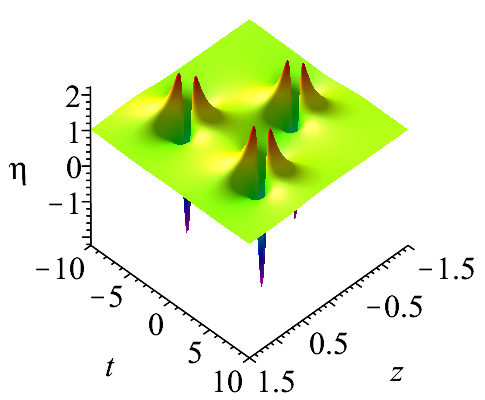}}
\caption{(Color online) The triangular structure of the second-order rogue wave solutions $(E,p,\eta)$ of the GNLS-MB system when $d=1, b=0, \omega=\frac{1}{2}, \tau=\frac{1}{2}, J_0=0, J_1=100$.}\label{fig:2r2}
\end{figure}

For $n=2$, the Eq. \eqref{nrw} in Theorem $2$ can give the second-order rogue wave solutions of the GNLS-MB system. when $J_0=J_1=0$, we can get the fundamental pattern which is plotted in Fig. \ref{fig:2r1}, the maximum amplitude of $|E^{[2]}_r|$ is five times as high as the background plane \cite{bib:Li}. When $J_0=0,\, J_1=100$, the fundamental pattern can be split into three first-order rogue waves (triangular structure \cite{bib:tri}), which is shown in Fig. \ref{fig:2r2}.

\begin{figure}[!htbp]
\centering
\subfigure{\includegraphics[height=4cm,width=5cm]{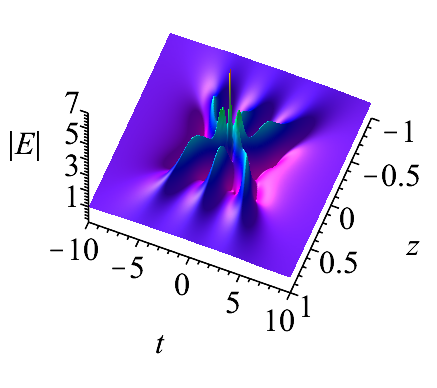}}
\subfigure{\includegraphics[height=4cm,width=5cm]{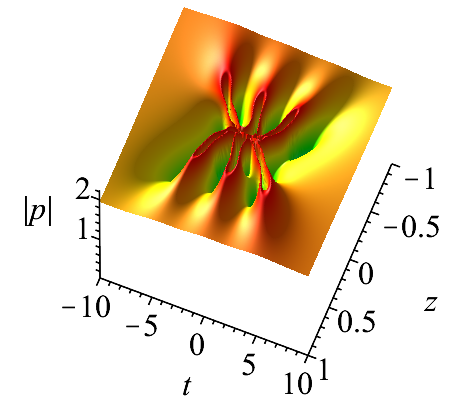}}
\subfigure{\includegraphics[height=4cm,width=5cm]{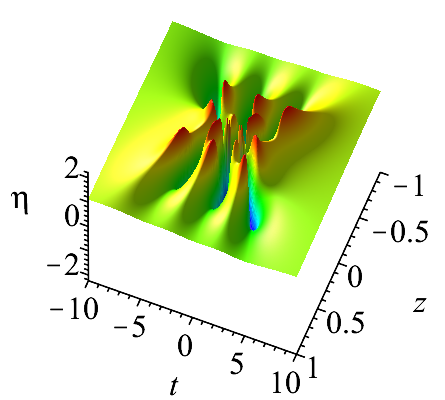}}
\caption{(Color online) The fundamental pattern of the third-order rogue wave solutions $(E,\, p,\, \eta)$ of the GNLS-MB system when $d=1,\, b=0,\, \omega=\frac{1}{2},\, \tau=\frac{1}{2},\, J_0=0,\, J_1=0,\, J_2=0$.}\label{fig:3r1}
\end{figure}

\begin{figure}[!htbp]
\centering
\subfigure{\includegraphics[height=4cm,width=5cm]{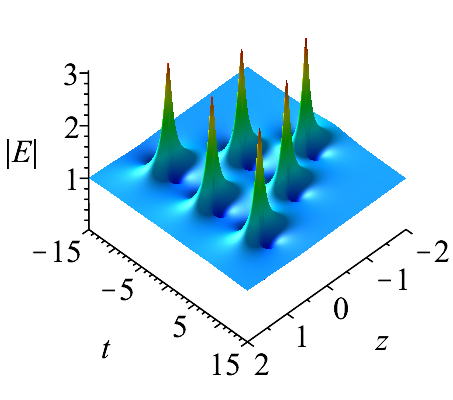}}
\subfigure{\includegraphics[height=4cm,width=5cm]{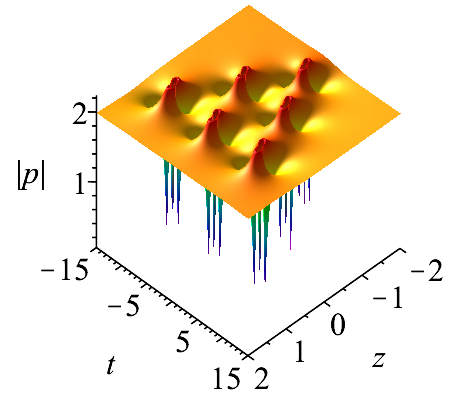}}
\subfigure{\includegraphics[height=4cm,width=5cm]{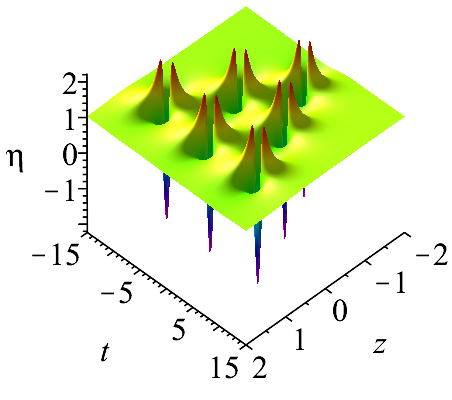}}
\caption{(Color online) The triangular structure of the third-order rogue wave solutions $(E,\, p,\, \eta)$ of the GNLS-MB system when $d=1,\, b=0,\, \omega=\frac{1}{2},\, \tau=\frac{1}{2},\, J_0=0,\, J_1=100,\, J_2=0$.}\label{fig:3r2}
\end{figure}

\begin{figure}[!htbp]
\centering
\subfigure{\includegraphics[height=4cm,width=5cm]{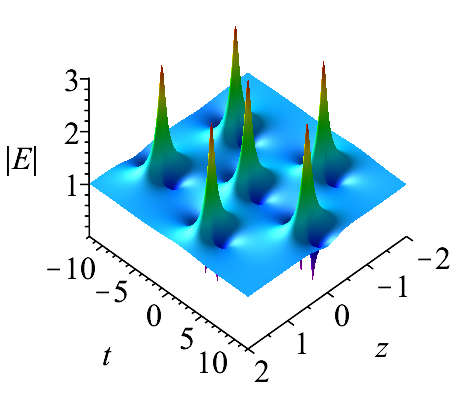}}
\subfigure{\includegraphics[height=4cm,width=5cm]{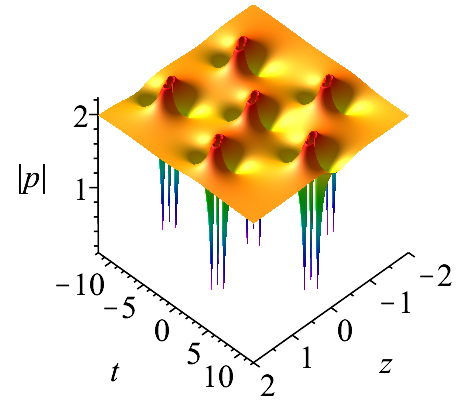}}
\subfigure{\includegraphics[height=4cm,width=5cm]{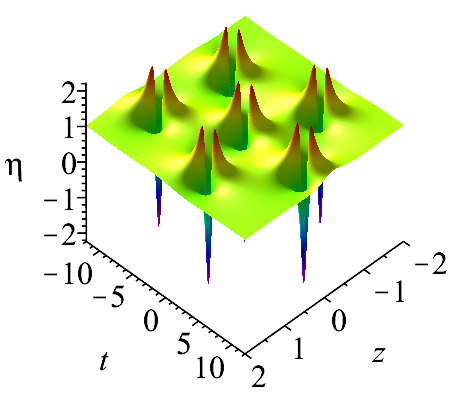}}
\caption{(Color online) The ring structure of the third-order rogue wave solutions $(E,\, p,\, \eta)$ of the GNLS-MB system when $d=1,\, b=0,\, \omega=\frac{1}{2},\, \tau=\frac{1}{2},\, J_0=0,\, J_1=0,\, J_2=5000$.}\label{fig:3r3}
\end{figure}

For $n=3$, the Eq. \eqref{nrw} can give the third-order rogue wave solutions of the GNLS-MB system. When $J_0=0,\, J_1=0,\, J_2=0$, we can get the fundamental pattern as plotted in Fig. \ref{fig:3r1}, the maximum amplitude of $|E^{[3]}_r|$ is seven times the height of the background plane. When $J_0=0,\, J_1=100,\, J_2=0$, the fundamental pattern is split into six first-order rogue waves (triangular structure), which is shown in Fig. \ref{fig:3r2}. But when $J_0=0,\, J_1=0,\, J_2=5000$, we get the ring structure which is also made up of six first-order rogue waves as in Fig. \ref{fig:3r3}.  Note that, the ring structure and the triangular structure have the same profile when $n=2$.

So far, we have obtained three types of structures: fundamental pattern, triangular structure and ring structure. According to the conclusions and decomposition rule of the higher-order rogue waves in Ref. \cite{bib:Li,bib:ring}, we know that higher-order rogue waves are generally the combination of the above three types. The specific application of these conclusions will be given in the following. The parameters $d,\, b,\, \omega,\, \tau$ take the same values as above.

\begin{figure}[!htbp]
\centering
\subfigure{\includegraphics[height=4cm,width=5cm]{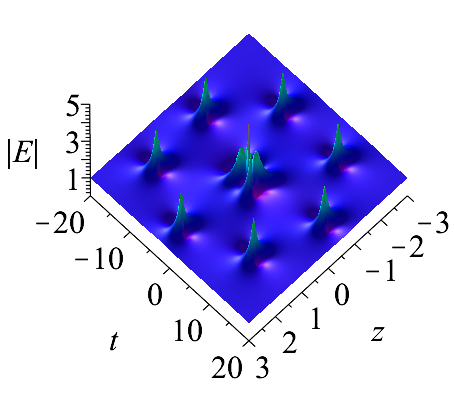}}
\subfigure{\includegraphics[height=4cm,width=5cm]{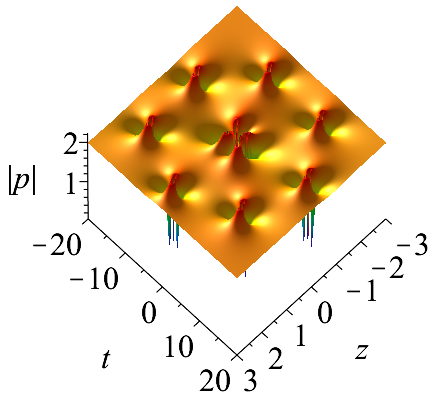}}
\subfigure{\includegraphics[height=4cm,width=5cm]{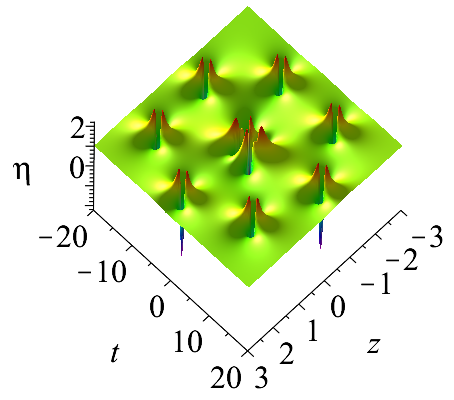}}
\caption{(Color online) The ring structure of the fourth-order rogue wave solutions $(E,\, p,\, \eta)$ of the GNLS-MB system when $d=1,\, b=0,\, \omega=\frac{1}{2},\, \tau=\frac{1}{2},\, J_0=0,\, J_1=0,\, J_2=0,\, J_3=10^6$.}\label{fig:4r3}
\end{figure}

\begin{figure}[!htbp]
\centering
\subfigure{\includegraphics[height=5cm,width=5cm]{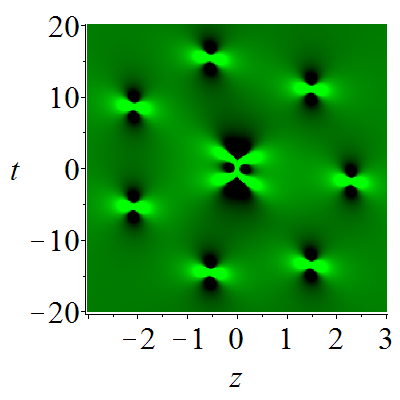}}
\subfigure{\includegraphics[height=5cm,width=5cm]{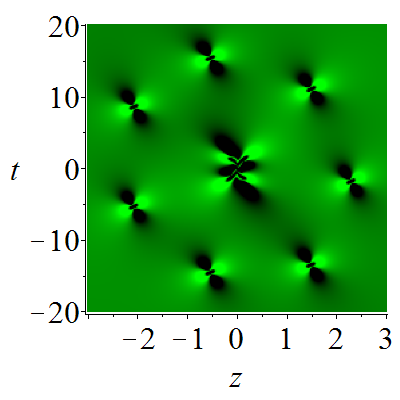}}
\subfigure{\includegraphics[height=5cm,width=5cm]{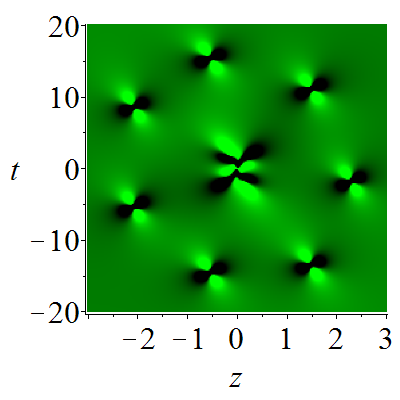}}
\caption{(Color online) Density  plots of the wave amplitudes for $(|E^{[4]}_r|,\, |p^{[4]}_r|,\, \eta^{[4]}_r)$ when $d=1,\, b=0,\, \omega=\frac{1}{2},\, \tau=\frac{1}{2},\, J_0=0,\, J_1=0,\, J_2=0,\, J_3=10^6$. They have a fundamental pattern in a ring.}\label{fig:4r31}
\end{figure}

\begin{figure}[!htbp]
\centering
\subfigure{\includegraphics[height=5cm,width=5cm]{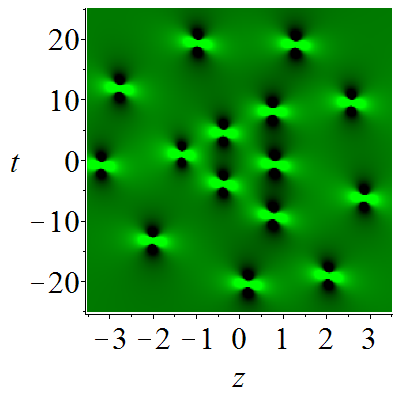}}
\subfigure{\includegraphics[height=5cm,width=5cm]{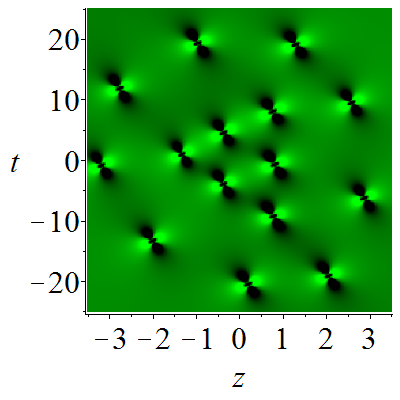}}
\subfigure{\includegraphics[height=5cm,width=5cm]{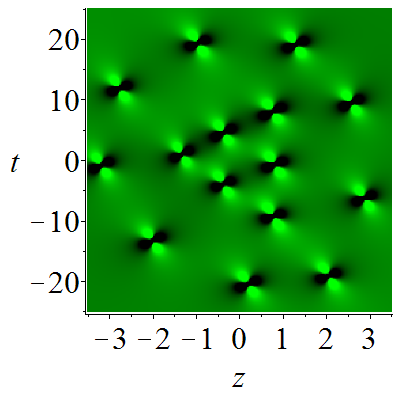}}
\caption{(Color online) Density plots of the wave amplitudes for $(|E^{[5]}_r|,\, |p^{[5]}_r|,\, \eta^{[5]}_r)$ when $d=1,\, b=0,\, \omega=\frac{1}{2},\, \tau=\frac{1}{2},\, J_0=0,\, J_1=100,\, J_2=0,\, J_3=0,\, J_4=10^{8}$. They have a triangular structure in  a ring.}\label{fig:5r41}
\end{figure}

\begin{figure}[!htbp]
\centering
\subfigure{\includegraphics[height=5cm,width=5cm]{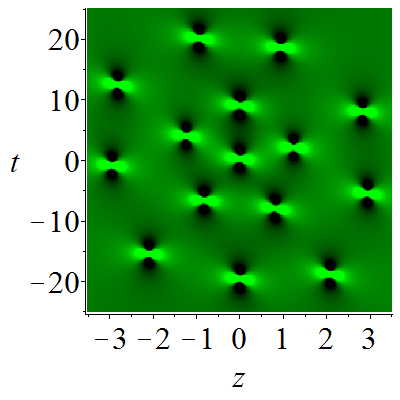}}
\subfigure{\includegraphics[height=5cm,width=5cm]{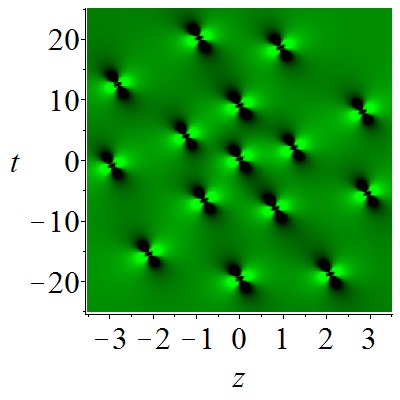}}
\subfigure{\includegraphics[height=5cm,width=5cm]{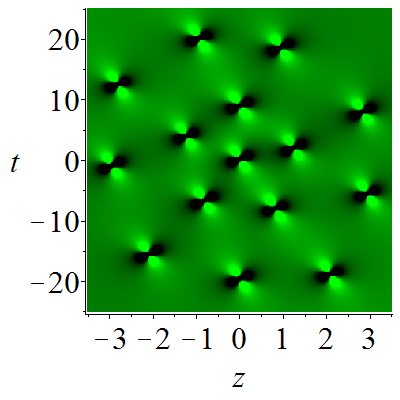}}
\caption{(Color online) Density  plots of the wave amplitudes for $(|E^{[5]}_r|,\, |p^{[5]}_r|,\, \eta^{[5]}_r)$ when $d=1,\, b=0, \omega=\frac{1}{2},\, \tau=\frac{1}{2},\, J_0=0,\, J_1=0,\, J_2=5000,\, J_3=0,\, J_4=10^{8}$. They have a ring structure in the outer ring, the outer ring is made up of nine first-order rogue, and the inner ring has five first-order rogue waves.}\label{fig:5r51}
\end{figure}

\begin{figure}[!htbp]
\centering
\subfigure{\includegraphics[height=5cm,width=5cm]{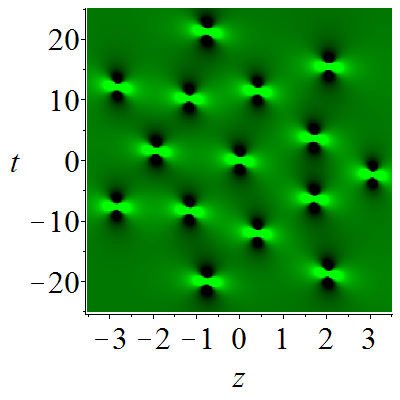}}
\subfigure{\includegraphics[height=5cm,width=5cm]{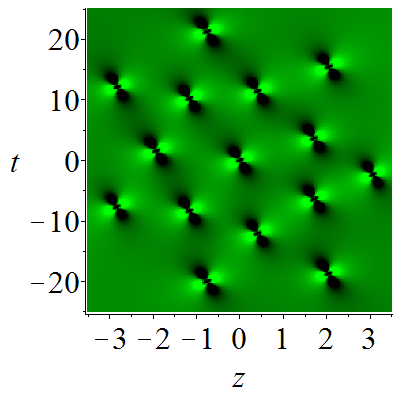}}
\subfigure{\includegraphics[height=5cm,width=5cm]{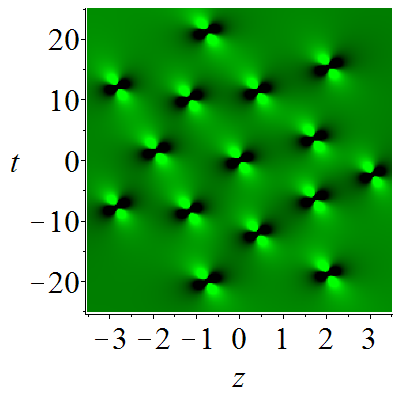}}
\caption{(Color online) Density  plots of the wave amplitudes of $(|E^{[5]}_r|,\, |p^{[5]}_r|,\, \eta^{[5]}_r)$ when $d=1,\, b=0,\, \omega=\frac{1}{2},\, \tau=\frac{1}{2},\, J_0=0,\, J_1=0,\, J_2=0,\, J_3=10^6,\, J_4=0$. They have two concentric rings, each of them consists of seven first-order rogue waves.}\label{fig:5r61}
\end{figure}

\begin{figure}[!htbp]
\centering
\subfigure{\includegraphics[height=5cm,width=5cm]{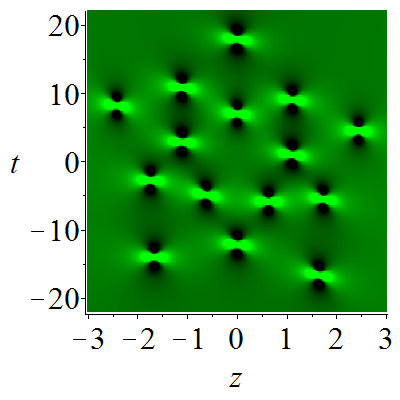}}
\subfigure{\includegraphics[height=5cm,width=5cm]{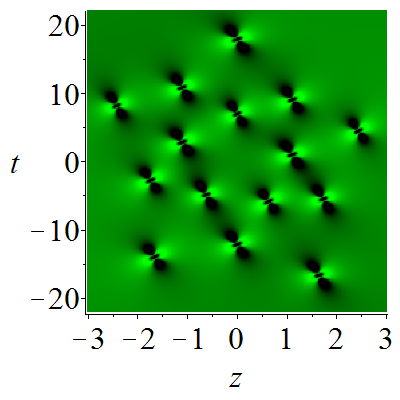}}
\subfigure{\includegraphics[height=5cm,width=5cm]{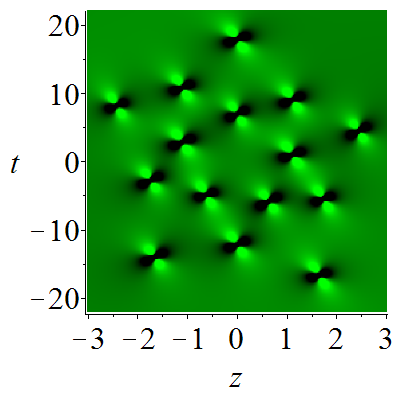}}
\caption{(Color online) Density  plots of the wave amplitudes of $(|E^{[5]}_r|,\, |p^{[5]}_r|,\, \eta^{[5]}_r)$ when $d=1,\, b=0,\, \omega=\frac{1}{2},\, \tau=\frac{1}{2},\, J_0=0,\, J_1=0,\, J_2=5000,\, J_3=0,\, J_4=0$. They have three concentric rings, each of them consists of five first-order rogue waves.}\label{fig:5r71}
\end{figure}

For the $n$th-order rogue waves in the Eq. \eqref{nrw}, it will result in the fundamental pattern if the parameters $J_j$ are all zero, then the fundamental pattern can split into triangular structure when $J_j=0$ except $J_1$. The parameter $J_{n-1}$ determines the ring structure when $n \geq 3$, $J_{n-1} \gg 1$ and other parameters $J_j$ are all zero. The ring structure consists of a fundamental pattern of $(n-2)$th-order rogue waves located in the center and $2n-1$ first-order rogue waves located on the outer circle. As Fig. \ref{fig:4r3} when $n=4$ with parameters $d=1,\, b=0,\, \omega=\frac{1}{2},\, \tau=\frac{1}{2},\, J_0=0,\, J_1=0,\, J_2=0,\, J_3=10^6$,  their corresponding density plots are shown in Fig. \ref{fig:4r31}. The fundamental pattern of $(n-2)$th-order rogue waves in the center can  keep on splitting into triangular structure or ring structure by choosing the parameters $J_j (j=0,\, 1,\, \cdots,\, n-3)$. And the process can go on indefinitely until they completely decompose. For example, we can get the ring structure of the fifth-order rogue waves by setting $J_4 \neq 0$. In this case: if we assume $J_1 \neq 0$, the inner third-order rogue waves can split into triangular structure (Fig. \ref{fig:5r41}); if we set $J_2 \neq 0$, the inner can split into a ring structure (Fig. \ref{fig:5r51}). Other parameters in addition to $J_{n-1}$ can also generate rings which are different from the above. For example, if and only if the parameter $J_{n-2} \neq 0$, the $n$th-order rogue waves have a double-ring structure, which has two rings with $2n-3$ first-order rogue waves for each ring and a fundamental pattern of $(n-4)$th-order rogue waves (Fig. \ref{fig:5r61}). And also if and only if $J_{n-3} \neq 0$, we can get a decomposition with three rings (Fig. \ref{fig:5r71}). We can also create more composite structures by selecting different combination of parameters appropriately. Here we will no longer show them in detail.

In fact, $J_0$ has no influence on the type of the $n$th-order rogue waves. It  can just adjust the position of the $n$th-order rogue waves on the background plane \cite{bib:zhang}. Next, we will discuss the nonlinear superposition of the rogue wave and breather solutions. Under the condition $C_1=C_2=1,\, \lambda_{2k-1}=-\frac{b}{2}+i\beta_{2k-1}$, substituting eigenfunctions in the Eq. \eqref{EF} into the Eq. \eqref{nsou}, we can get the $n$th-order breather solutions. We know that the $n$th-order rogue waves could be generated by assuming $\beta_{2k-1} \rightarrow d (k=1,\, 2,\, \cdots, n)$. When $n \geq 2$, if we only take the limit for $\beta_{2k-1} (k=1,\, 2,\, \cdots,\, n_r), \ 1 \le n_r < n$, we can construct new solutions which are the nonlinear superposition of $n_r$th-order rogue waves and $(n-n_r)$th-order breather solutions. In order to separate the $n_r$th-order rogue waves from the $(n-n_r)$th-order breather solutions, we can modify $C_1$ and $C_2$ by substituting the Eq. \eqref{rwC} into eigenfunctions in the Eq. \eqref{EF} only when $k=1,\, 2,\, \cdots,\, n_r$. Note that, we take $C_1=C_2=1$ in the Eq. \eqref{EF} when $k=n_r+1,\, \cdots,\, n$.

\begin{figure}[!htbp]
\centering
\subfigure{\includegraphics[height=4cm,width=5cm]{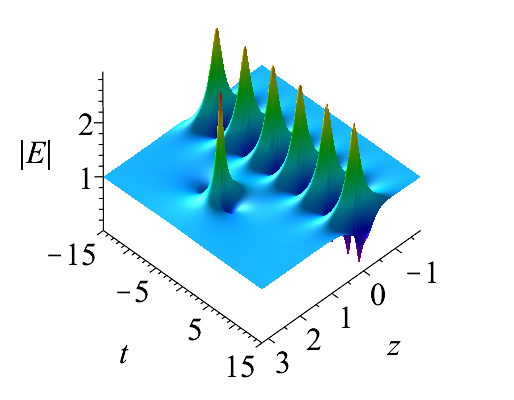}}
\subfigure{\includegraphics[height=4cm,width=5cm]{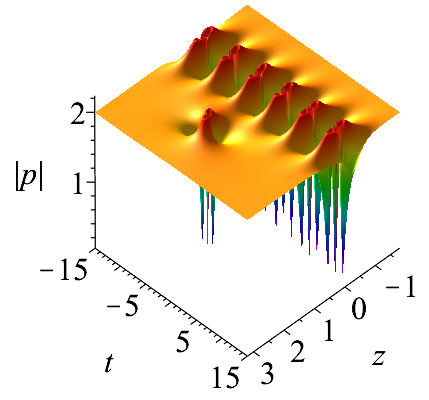}}
\subfigure{\includegraphics[height=4cm,width=5cm]{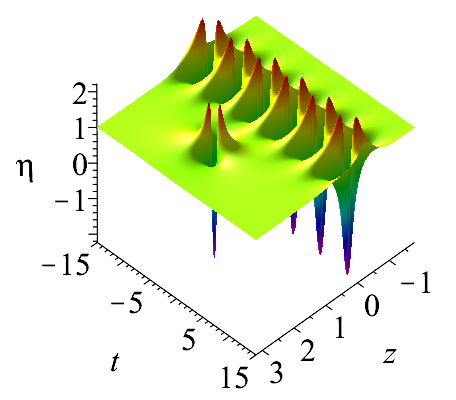}}
\caption{(Color online) The completely isolated form of the solutions which is the nonlinear superposition of the first-order rogue waves and first-order breather solutions, when $d=1,\, b=0,\, \omega=\frac{1}{2},\, \tau=\frac{1}{2},\, \beta_3=\frac{4}{5},\, J_0=-10i$.}\label{fig:rb2}
\end{figure}

\begin{figure}[!htbp]
\centering
\subfigure{\includegraphics[height=4cm,width=5cm]{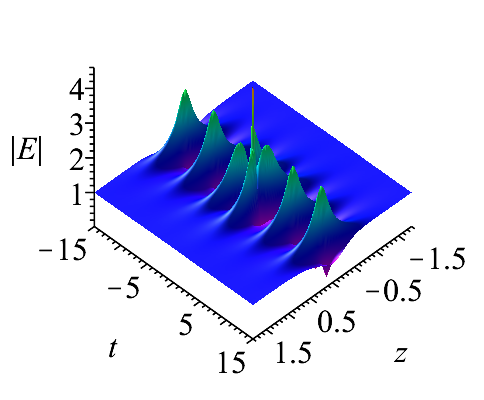}}
\subfigure{\includegraphics[height=4cm,width=5cm]{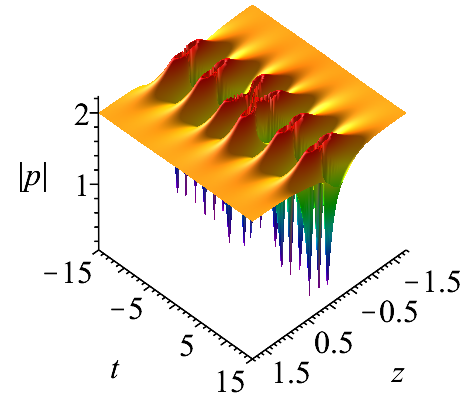}}
\subfigure{\includegraphics[height=4cm,width=5cm]{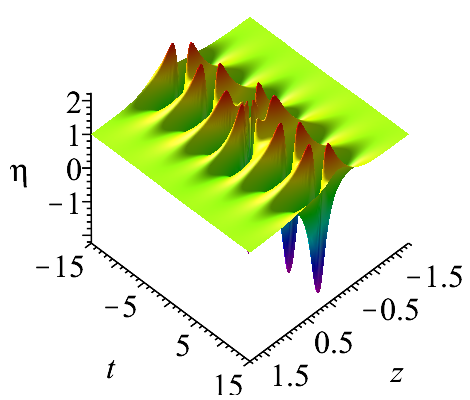}}
\caption{(Color online) The form shown as a first-order breather with a central fundamental pattern of second-order rogue waves, when $d=1,\, b=0,\, \omega=\frac{1}{2},\, \tau=\frac{1}{2},\, \beta_3=\frac{4}{5},\, J_0=0$.}\label{fig:rb1}
\end{figure}

Due to the tediousness of the solutions, we give the dynamical evolution of the results. For $n=2,\, n_r=1$, when the two components are separated completely, the solutions appear as a first-order rogue wave and a first-order breather solution as plotted in Fig. \ref{fig:rb2}, which are similar to Fig. 3(a) in Ref. \cite{bib:hun}. However, when the two components are nonlinearly overlaid, the solutions appear as a first-order breather with a central fundamental pattern of a second-order rogue wave as plotted in Fig. \ref{fig:rb1}, which are similar to Fig. 3(b) in Ref. \cite{bib:hun}. Comparing Fig. \ref{fig:rb2} with Fig. \ref{fig:rb1}, we observe that the three first-order peaks merge into a second-order peak, in other words, a second-order peak splits into three first-order peaks. These hybrid solutions give the interaction between the rogue wave and breather solutions.

\section{conclusion} \label{sec:6}

In this article, we derived the Darboux transformation of the GNLS-MB system which describes the ultrashort pulse propagation in the resonant erbium-doped nonlinear optical fiber with higher-order effects, and constructed the determinant representation of the $n$-fold Darboux transformation in Theorem 1. We established the determinant representation of the solutions $(E^{[n]},\, p^{[n]},\, \eta^{[n]})$ generated from the known trivial seed solutions $(E,\, p,\, \eta)$ in the Eq. \eqref{nsou}. Under the reduction conditions $\lambda_{2k}=\lambda_{2k-1}^*$ and $f_{2k}=\big(\begin{smallmatrix} -f_{2k-12}^* \\ f_{2k-11}^* \end{smallmatrix}\big)$, we got the breather solutions from the nonzero periodic seed solutions. Using the method of limit technique and Taylor expansions, we constructed the determinant representation of the $n$th-order rogue waves in Theorem 2. The rogue waves show interesting characteristics that might attract physicists in experiments to observe them, in contrast with the common bright rogue wave $E$, the dark rogue waves for $p$ and $\eta$ have two (or more) dominant down peaks in their profiles, and there is an upper ring in the profile of $p$.

By  analyzing higher-order terms, we found that $\tau$ has an effect on the periodic of the breather solutions except Ma breather solutions. By calculating, we knew $p$ and $\eta$ had the same locations in connection with $\tau$  and that $\tau$ did not affect the values of their extremes.

In the last section of this article, we also gave the three basic types of the $n$th-order rogue waves, and combination structures could be obtained by choosing proper parameters. At the end of the last section, we introduce the hybrid solutions which are the nonlinear superposition of the rogue wave and breather solutions. These solutions display the interaction between the rogue wave and breather solutions, which could help us  to understand the generation of rogue waves better. Moreover, the exploration for other coupled systems with more complex higher-order optical effects can be done in our future work.

%%%%%%%%%%%%%%%%%%%%%%%%%%%%%%%%%%%%%%%%%%%%%%%%%%%%%%%%%%%%%%%%%%%%%%%%%%%%%%%%
%%%%%%%%%%%%%%%%%%%%%%%%%%%%%%%%%%%%%%%%%%%%%%%%%%%%%%%%%%%%%%%%%%%%%%%%%%%%%%%%5
%%%%%%%%%%%%%%%%%%%%%%%%%%%%%%%%%%%%%%%%%%%%%%%%%%%%%%%%%%%%5

{\bf Acknowledgments}

{\noindent \small
Jingsong He is supported by the National Natural Science Foundation of China under Grant No. 11271210, K. C. Wong Magna Fund in Ningbo University.
 Chuanzhong Li is supported by the National Natural Science Foundation of China under Grant No. 11201251, 11571192,
 the Zhejiang Provincial Natural Science Foundation under Grant No. LY15A010004, LY12A01007, the Natural Science Foundation of Ningbo under Grant No. 2015A610157. }

%%%%%%%%%%%%%%%%%%%%%%%%%%%%%%%%%%%%%%%%%%%

%%%%%%%%%%%%%%%%%%%%%%%%%%%%%%%%%%%%%%%%%%%
%\begin{figure}[htbp]
 % \centering\mbox{}\hspace{0cm}
 % \subfigure[$a  = 0 ,c = 1,\delta  = \frac{1}{{100}},\alpha  =\frac{1}{{100}}$]{\label{fig:11}\includegraphics[width=0.45\textwidth]{1r11jb.jpg}}
 % \subfigure[$a  = 0 ,c = 1,\delta  = \frac{1}{{100}},\alpha  =\frac{1}{{100}}$]{\label{fig:12}\includegraphics[width=0.45\textwidth]{2r11jb.jpg}}
 % \subfigure[$a  = 0 ,c = 1,\delta  = \frac{1}{{100}},\alpha  =\frac{1}{{100}}$]{\label{fig:21}\includegraphics[width=0.45\textwidth]{3r11jb.jpg}}
 % \subfigure[$a  = 0 ,c = 1,\delta  = \frac{1}{{100}},\alpha  =\frac{1}{{100}}$]{\label{fig:22}\includegraphics[width=0.45\textwidth]{4r11jb.jpg}}
 % \caption{The rogue wave solutions(the Fundamental pattern) up to order 4 with zero shifts.}
%\label{fig:1}
%\end{figure}

\newpage

%\begin{figure}[!htbp]
%\centering
%\subfigure{\includegraphics[height=5cm,width=5cm]{4-r-4E1.png}}
%\subfigure{\includegraphics[height=5cm,width=5cm]{4-r-4p1.png}}
%\subfigure{\includegraphics[height=5cm,width=5cm]{4-r-4eta1.png}}
%\caption{(Color online) The ring structure of the third-order rogue wave solutions $(E,p,\eta)$ of the GNLS-MB system when $d=1, b=0, \omega=\frac{1}{2}, %\tau=\frac{1}{2}, J_0=0, J_1=100, J_2=0, J_3=10^6$.}\label{fig:4r41}
%\end{figure}

%\begin{figure}[!htbp]
%\begin{minipage}{0.3\linewidth}\centering
%\includegraphics[height=4cm,width=5cm]{pp1.png}
%\caption{(Color online) The first-order breather solutions $(E,p,\eta)$ of the GNLS-MB system when $d=1, b=0, \omega=\frac{1}{2}, \tau=\frac{1}{2}, \beta_3=\frac{4}{5}$.}\label{fig:rb1}
%\end{minipage}
%\begin{minipage}{0.3\linewidth}\centering
%\includegraphics[height=4cm,width=5cm]{pp2.png}
%\caption{(Color online) The first-order $(E,p,\eta)$ of the GNLS-MB system when $\tau=\frac{1}{2}$.}\label{fig:rb1}
%\end{minipage}
%\begin{minipage}{0.3\linewidth}\centering
%\includegraphics[height=4cm,width=5cm]{pp3.png}
%\caption{(Color online) The first-order $(E,p,\eta)$ of the GNLS-MB system when $\tau=\frac{1}{2}$.}\label{fig:rb1}
%\end{minipage}
%\end{figure}

\end{document}